\documentclass[12pt]{article}
\usepackage{epsf}

\renewcommand{\bar}{\overline}

\def\ru1{\rule[-0.4truecm]{0mm}{1truecm}}

\renewcommand{\bar}[1]{\overline{#1}}
\newcommand{\M}{{\cal M}}
\newcommand{\VEV}[1]{\left\langle{#1}\right\rangle}
\newcommand{\etal}{{\em et al.}}
\newcommand{\ie}{{\it i.e.}}
\newcommand{\eg}{{\it e.g.}}
\newcommand{\half}{{1\over 2}}

\newcommand{\ket}[1]{\vert\,{#1}\rangle}

\renewcommand{\bar}{\overline}

\newcommand{\hsim}[1]{\ \ \mathrel{\rlap{\lower-4pt\hbox{$\sim$}}
                    \hskip-3pt\hbox{$#1$}}\,}

\begin{document}
\begin{flushright}
{\small
SLAC--PUB--8770\\
February 2001}
\end{flushright}

\bigskip\bigskip
\begin{center}
{{\bf\LARGE
Hadronic Light-Front Wavefunctions\\[1ex]
 and QCD Phenomenology}\footnote{Work supported by
Department of Energy contract  DE--AC03--76SF00515.}}

\bigskip
\bigskip
Stanley J. Brodsky\\
{{\sl  Stanford Linear Accelerator Center \\
Stanford, California 94309} \\
E-mail: sjbth@slac.stanford.edu}

\end{center}

\vfill

\begin{center}
{Invited lectures presented at the \\
Fifth International Workshop On Particle Physics Phenomenology \\
   Chi-Pen, Taitung, Taiwan\\
         8--11 November 2000}
\end{center}

\vfill\eject

\begin{center}
{\bf\large Abstract }
\end{center}

A fundamental goal in QCD is to understand
the non-perturbative structure
of hadrons at the amplitude level---not just the single-particle flavor,
momentum, and helicity distributions of the quark constituents,  but also
the multi-quark, gluonic, and hidden-color correlations intrinsic to
hadronic and nuclear wavefunctions.
A natural calculus for describing
the bound-state structure of relativistic composite systems in quantum
field theory is the light-front Fock expansion which encodes the
properties of a hadrons in terms of a set of frame-independent $n-$particle
wavefunctions.
Light-front quantization in the doubly-transverse
light-cone gauge has a number of remarkable advantages,
including explicit unitarity, a physical Fock
expansion, the absence of ghost degrees of freedom,
and the decoupling properties needed to prove
factorization theorems in high momentum transfer inclusive and exclusive
reactions.  A number of applications are discussed in these lectures,
including semileptonic $B$ decays, two-photon exclusive reactions, and
deeply virtual Compton
scattering. The relation of the intrinsic sea to the
light-front wavefunctions is  discussed.  A
new type of jet production reaction, ``self-resolving diffractive
interactions" can provide direct information on the light-front
wavefunctions of hadrons in terms of their quark and gluon degrees of
freedom as well as the composition of nuclei in terms of their nucleon
and mesonic degrees of freedom.

\bigskip\bigskip

\renewcommand{\baselinestretch}{1.2}
\normalsize

\section{Introduction}

In principle,  quantum chromodynamics provides a fundamental description
of hadron and nuclear physics in terms of quark and gluon
degrees of freedom.  QCD has been developed and successfully tested
extensively, particularly in inclusive and
exclusive processes involving collisions at large momentum transfer
where factorization theorems and the smallness of the QCD effective
coupling allow perturbative predictions.  However, despite its empirical
successes, many fundamental questions about QCD have not been resolved.
These include a rigorous proof of color confinement,
the behavior of the QCD
coupling at small momenta,
a rigorous description of the structure of hadrons in terms of their
quark and gluon degrees of freedom,  the problem of asymptotic $n!$
growth of the perturbation theory (renormalon phenomena),  the
nature of the pomeron and Reggeons, the nature of shadowing
and anti-shadowing in nuclear collisions, the apparent conflict between
QCD vacuum structure and the small size of the cosmological constant, and
the problems of scale and scheme ambiguities in perturbative QCD
expansion.

In these lectures I will focus on one of the central questions in QCD---the
non-perturbative
description of the
proton and other hadrons as composites of confined, relativistic
quark and gluon quanta.  The goal is a
frame-independent, quantum-mechanical representation of hadrons at the
amplitude level capable of encoding multi-quark, hidden-color and gluon
momentum, helicity, and flavor correlations in the form of universal
process-independent hadron wavefunctions.  For example, the
measurement and interpretation of the basic parameters of the electroweak
theory and
$CP$ violation depends on an understanding of the dynamics and phase
structure of $B$ decays at the amplitude level.  As I will discuss in
these lectures, light-front quantization allows a unifying
wavefunction representation of non-perturbative hadron dynamics in QCD.
Remarkably, it is now possible to measure the wavefunctions of a
relativistic hadron by diffractively dissociating it into jets
whose momentum distribution is correlated with the valence quarks'
momenta  \cite{Ashery:1999nq,Bertsch:1981py,Frankfurt:1993it,Frankfurt:2000tq}.
It is also particularly important to understand the shape of the
gauge- and process-independent meson and baryon valence-quark
distribution amplitudes \cite{Lepage:1980fj}
$\phi_M(x,Q)$, and
$\phi_B(x_i,Q)$. These quantities specify how a hadron shares its
longitudinal momentum among its valence quarks; they control virtually
all exclusive processes involving a hard scale
$Q$, including form factors, Compton scattering and photoproduction
at large momentum transfer, as well as the decay of a heavy hadron into
specific final states \cite{Beneke:1999br,Keum:2000ph}.

What do we mean by a hadronic wavefunction, and what are its degrees of
freedom?  In the chiral effective Lagrangian approximation, a baryon can be
represented as a classical soliton solution as in the Skyrme model.  On
the other hand, bag models, the observed baryon spectroscopy, and
magnetic moment phenomenology suggests that baryons are composites of
three ``constituent" quarks or perhaps a quark bound to an effective
spin-0 or spin-1 diquark.  The cloudy bag model \cite{Signal:1987gz} takes
into account the effect of meson-baryon fluctuations such as $n \pi^+$ and
$\Lambda K^+$ in the proton.  This in turn implies the existence of sea
quarks with an $\bar u(x) \ne \bar d(x)$ and $s(x) \ne \bar s(x)$
asymmetries in the momentum and spin distributions of the quark sea.

The $n$-particle Schr\"odinger momentum space
wavefunction $\psi_N({\vec p}_i)$ of a composite system is the
projection of the exact eigenstate of the equal-time Hamiltonian on the
$n$-particle states of the non-interacting Hamiltonian, the Fock basis.
It represents the amplitude for finding the constituents with
three-momentum ${\vec p}_i$, orbital angular momentum, and spin, subject
to three-momentum conservation and angular momentum sum rules.  The
constituents are on their mass shell, $E_i = \sqrt{{\vec p}^2_i +
m^2_i}$ but do not conserve energy $\sum^n_{i = 1}E_i > E = \sqrt{{\vec
p}^2 + M^2}$.  However, in a relativistic quantum theory, a bound-state
cannot be represented as a state with a fixed number of constituents.
For example, the existence of gluons which propagate between the valence
quarks necessarily implies that the hadron wavefunction must describe
states with an arbitrary number of gluons.  Thus a hadronic wavefunction
must describe fluctuations in particle number $n$, as well as momenta and
spin.  One has to take into account fluctuations in the wavefunction
which allow for any number of sea quarks, as long as the total quantum
numbers of the constituents are compatible with the overall quantum
numbers of the baryon.

Gluon exchange between quarks of different nucleons in a nucleus will
leave the nucleons in an exited octet state of $SU(3)_C$, thus requiring
the existence of ``hidden color" configurations in a nuclear
wavefunction.  What is worse
is that the description of the state at a given time $t_0$ depends not
only on the choice of gauge but also on the Lorentz frame of the observer
since two observers will differ on their definition of simultaneity.

A number of non-perturbative formalisms have been developed which in
principle could provide the exact spectra and wavefunctions of bound
states in relativistic quantum field theory, including QCD.  These include
Euclidean lattice gauge theory, the covariant
Bethe-Salpeter bound state equation, and
Hamiltonian methods based on equal-time or light-front quantization.
Lattice gauge theory, particularly in the heavy quark approximation, has
been an enormously helpful guide to the hadronic and glueball spectrum
of QCD; however, progress in evaluating hadron wavefunctions has been so far
limited to the evaluation of certain moments of
hadronic matrix elements \cite{Martinelli:1987si}.

The
Bethe-Salpeter formalism in principle provides
a rigorous and systematic, renormalizable, and explicitly
covariant treatment of bound states of two interacting fields.  There has
been recent progress applying model forms of Dyson-Schwinger equations to
the Bethe-Salpeter formalism in order to incorporate the effects of
dynamical symmetry breaking in QCD \cite{Hecht:2000xa}. However, the actual
interaction kernel in the bound state equation and the evaluation of current
matrix elements requires a sum over an infinite number of irreducible
graphs; the use of a single kernel (the ladder approximation) breaks gauge
invariance, crossing symmetry, and has an improper heavy quark limit.  It
is also intrinsically difficult to apply the Bethe Salpeter to bound
states of three or more constituents.

There has been progress applying
equal-time Hamiltonian methods to QCD in Coulomb
gauge \cite{Robertson:1999va}; however, the renormalization procedure in
this formalism is not well-defined and the results are typically
frame-dependent.  The evaluation of current matrix elements in the
equal-time formalism is especially problematic since one must include
contributions to the current from vacuum excitations.  For similar reasons,
the expansion of an eigenstate on the equal-time Fock basis of free quanta
is not well-defined.

In the light-front quantization,
one takes the light-cone time variable
$ t + z/c$ as the evolution parameter instead of ordinary time $t$.
(The $\hat z$ direction is an arbitrary reference direction.)
The method is often called
``light-front" quantization rather than ``light-cone" quantization since
the equation
$x^+ = \tau = 0$ defines a hyperplane corresponding to a
light-front.
The light-front fixes the initial boundary conditions
of a composite system as its constituents are
intercepted by a light-wave evaluated at a specific value
of
$x^+ = t + z/c$.  In contrast, determining an atomic wavefunction at a
given instant
$t = t_0$ requires measuring the simultaneous scattering of $Z$ photons on
the
$Z$ electrons.

An extensive review and guide
to the light-front quantization literature can be
found in Ref. \cite{Brodsky:1998de}.  I will
use here the notation $A^\mu = (A^+, A-, A_\perp),$ where $A^\pm = A^0
\pm A^z$ and the metric $A
\cdot B = {1\over 2 }( A^+ B^- + A^- B^+ ) - A_\perp \cdot B_\perp.$

The origins of the light-front quantization method can be traced to
Dirac \cite{Dirac:1949cp} who noted that 7 out of the 10 Poincare'
generators, including a Lorentz boost $K_3$, are kinematical
(interaction-independent) when one quantizes a theory
at fixed light-cone time.  This in turn
leads to the remarkable property that the light-front wavefunctions of a
hadron are independent of the hadron's total momentum, whether it is at
rest or moving.  Thus once one has solved for the light-front
wavefunctions, one can compute hadron matrix elements of currents between
hadronic states of arbitrary momentum.  In
contrast, knowing the rest frame wavefunction at equal time, does not
determine the moving hadron's wavefunction.

Light-front wavefunctions are related to momentum-space Bethe-Salpeter
wavefunctions by integrating over the relative momenta
$k^- = k^0 - k^z$ since this projects out $x^+ =0.$ The light-front Fock
space is the eigenstates of the free light-front Hamiltonian; \ie, it is a
Hilbert space of non-interacting quarks and gluons, each of which
satisfy $k^2 = m^2$ and $k^- = {m^2 + k^2_\perp \over k^+} \ge 0.$
The light-front wavefunctions are the projections of the hadronic
eigenstate on the light-front Fock basis, the complete set of color singlet
states of the free Hamiltonian.  An essentially equivalent approach,
pioneered by Weinberg \cite{Weinberg:1966jm,Brodsky:1973kb}, is to evaluate
the equal-time theory from the perspective of an observer moving in the
negative $\hat z$ direction with arbitrarily large momentum $P_z \to
-\infty.$ The light-cone fraction $x = {k^+\over p^+}$ of a constituent
can be identified with the longitudinal momentum $x = {k^z\over P^z}$ in a
hadron moving with large momentum $P^z.$

It is convenient to define the invariant light-front
Hamiltonian: $ H^{QCD}_{LC} = P^+ P^- - {\vec P_\perp}^2$ where
$P^\pm = P^0 \pm P^z$.  The operator
$P^- = i {d\over d\tau}$ generates light-cone time translations.
The
$P^+$ and
$\vec P_\perp$ momentum operators are independent of
the interactions, and thus are conserved at all orders.
The eigen-spectrum of $ H^{QCD}_{LC}$ in principle gives the entire mass
squared spectrum of color-singlet hadron states in QCD, together with
their respective light-front wavefunctions.  For example, the
proton state satisfies:
$ H^{QCD}_{LC} \ket{\Psi_p} = M^2_p \ket{\Psi_p}$.
The projection of
the proton's eigensolution $\ket{\Psi_p}$ on the color-singlet
$B = 1$, $Q = 1$ eigenstates $\{\ket{n} \}$
of the free Hamiltonian $ H^{QCD}_{LC}(g = 0)$ gives the
light-front Fock expansion: \cite{Brodsky:1989pv}
\begin{eqnarray}
\left\vert \Psi_p; P^+, {\vec P_\perp}, \lambda \right> &=&
\sum_{n \ge 3,\lambda_i}  \int \Pi^{n}_{i=1}
{d^2k_{\perp i} dx_i \over \sqrt{x_i} 16 \pi^3} \cr
&& 16 \pi^3 \delta\left(1- \sum^n_j x_j\right) \delta^{(2)}
\left(\sum^n_\ell \vec k_{\perp \ell}\right) \cr
&&\left\vert n;
x_i P^+, x_i {\vec P_\perp} + {\vec k_{\perp i}}, \lambda_i\right >
\psi_{n/p}(x_i,{\vec k_{\perp i}},\lambda_i)
 . \nonumber
\end{eqnarray}

It is especially convenient to develop the light-front formalism in the
light-cone gauge $A^+ = A^0 + A^z = 0$.  In this gauge the $A^-$ field
becomes a dependent degree of freedom, and it can be eliminated from the
gauge theory Hamiltonian, with the addition of a set of
specific instantaneous light-cone time
interactions.  In fact in $QCD(1+1)$ theory, this instantaneous
interaction provides the confining linear $x^-$ interaction between
quarks.  In $3+1$ dimensions, the transverse field $A^\perp$ propagates
massless spin-one gluon quanta with polarization vectors \cite{Lepage:1980fj}
which satisfy both the gauge condition $\epsilon^+_\lambda = 0$ and the
Lorentz condition $k\cdot \epsilon= 0$.  Thus no extra condition on the
Hilbert space is required.

The light-front Fock
wavefunctions $\psi_{n/H}(x_i,{\vec k_{\perp i}},\lambda_i)$
thus interpolate between the hadron $H$ and
its quark and gluon degrees of freedom.
The light-cone momentum fractions of the constituents,
$x_i = k^+_i/P^+$ with $\sum^n_{i=1} x_i = 1,$ and the transverse
momenta ${\vec k_{\perp i}}$ with
$\sum^n_{i=1} {\vec k_{\perp i}} = {\vec 0_\perp}$ appear as
the momentum
coordinates of the light-front Fock wavefunctions.  A crucial feature is
the frame-independence of the light-front wavefunctions.  The $x_i$ and
$\vec k_{\perp i}$ are relative coordinates independent of the hadron's
momentum $P^\mu$.  The actual physical transverse momenta are
${\vec p_{\perp i}} = x_i {\vec P_\perp} + {\vec k_{\perp i}}.$
The $\lambda_i$ label the light-front spin $S^z$ projections of the
quarks and
gluons along the $z$ direction.  The physical gluon
polarization vectors
$\epsilon^\mu(k,\ \lambda = \pm 1)$ are specified in light-cone
gauge by the conditions $k \cdot \epsilon = 0,\ \eta \cdot \epsilon =
\epsilon^+ = 0.$
Each light-front Fock wavefunction satisfies conservation of the
$z$ projection of angular momentum:
$
J^z = \sum^n_{i=1} S^z_i + \sum^{n-1}_{j=1} l^z_j \ .
$
The sum over $S^z_i$
represents the contribution of the intrinsic spins of the $n$ Fock state
constituents.  The sum over orbital angular momenta
$l^z_j = -{\mathrm i} (k^1_j\frac{\partial}{\partial k^2_j}
-k^2_j\frac{\partial}{\partial k^1_j})$ derives from
the $n-1$ relative momenta.  This excludes the contribution to the
orbital angular momentum due to the motion of the center of mass, which
is not an intrinsic property of the hadron \cite{Brodsky:2001ii}.

The interaction Hamiltonian of QCD in light-cone gauge can be derived by
systematically applying the Dirac bracket method to identify the independent
fields \cite{Srivastava:2000cf}.
It contains the usual Dirac interactions
between the quarks and gluons, the three-point and four-point gluon
non-Abelian interactions plus instantaneous
light-front-time gluon exchange and quark exchange contributions
\begin{eqnarray}
{\cal H}_{int}&=&
  -g \,{{\bar\psi}}^{i}
\gamma^{\mu}{A_{\mu}}^{ij}{{\psi}}^{j}   \nonumber \\
&& +\frac{g}{2}\,
f^{abc} \,(\partial_{\mu}{A^{a}}_{\nu}-
\partial_{\nu}{A^{a}}_{\mu}) A^{b\mu} A^{c\nu} \nonumber \\
&& +\frac {g^2}{4}\,
f^{abc}f^{ade} {A_{b\mu}} {A^{d\mu}} A_{c\nu} A^{e\nu} \nonumber \\
&& - \frac{g^{2}}{ 2}\,\, {{\bar\psi}}^{i}
\gamma^{+}
\,(\gamma^{\perp'}{A_{\perp'}})^{ij}\,\frac{1}{i\partial_{-}} \,
(\gamma^{\perp} {A_{\perp}})^{jk}\,{\psi}^{k} \nonumber \\
&& -\frac{g^{2}}{ 2}\,{j^{+}}_{a}\, \frac {1}{(\partial_{-})^{2}}\,
{j^{+}}_{a}
\end{eqnarray}
where
\begin{equation}
{j^{+}}_{a}={{\bar\psi}}^{i}
\gamma^{+} ( {t_{a}})^{ij}{{\psi}}^{j}
+ f_{abc} (\partial_{-}
A_{b\mu}) A^{c\mu} \ .
\end{equation}

In
light-cone time-ordered perturbation theory, a Green's functions is
expanded as a power series in the interactions with light-front energy
denominators $\sum_{\rm initial} k^-_i - \sum_{\rm intermediate} k^-_i + i
\epsilon$ replacing the usual energy denominators.  [For a review
see Ref. \cite{Brodsky:1989pv}.]
In general each Feynman diagram with $n$
vertices corresponds to the sum of
$n!$ time-ordered contributions.  However, in light-cone-time-ordered
perturbation theory, only those few graphs where all $k^+_i
\ge 0$ survive.  In addition the form of the light-front kinetic energies
is rational: $k^- = {k^2_\perp + m^2 \over k^+}$, replacing the
nonanalytic
$k^0 = \sqrt{{\vec k}^2 + m^2}$ of equal-time theory.
Thus light-cone-time-ordered perturbation theory provides a viable
computational method where one can trace the physical evolution of a
process.  The integration measures are only three-dimensional $d^2k_\perp
dx$; in effect, the $k^-$ integral of the covariant perturbation theory
is performed automatically.

Alternatively, one derive Feynman rules for QCD in
light-cone gauge, thus allowing the use of standard covariant
computational tools and renormalization methods including dimensional
regularization.  Prem Srivastava and I \cite{Srivastava:2000cf} have
recently presented a systematic study of light-front-quantized gauge theory
in light-cone gauge using a Dyson-Wick S-matrix expansion based on
light-cone-time-ordered products.  The gluon propagator
has the form
\begin{equation}
\VEV{0|\,T({A^{a}}_{\mu}(x){A^{b}}_{\nu}(0))\,|0} ={{i\delta^{ab}} \over
{(2\pi)^{4}}} \int d^{4}k \;e^{-ik\cdot x} \; \; {D_{\mu\nu}(k)\over
{k^{2}+i\epsilon}}
\end{equation}
where we have defined
\begin{equation}
D_{\mu\nu}(k)= D_{\nu\mu}(k)= -g_{\mu\nu} + \frac
{n_{\mu}k_{\nu}+n_{\nu}k_{\mu}}{(n\cdot k)} - \frac {k^{2}} {(n\cdot
k)^{2}} \, n_{\mu}n_{\nu}.
\end{equation}
Here $n_{\mu}$ is a null
four-vector, gauge direction, whose components are chosen to be $\,
n_{\mu}={\delta_{\mu}}^{+}$, $\, n^{\mu}={\delta^{\mu}}_{-}$.  Note also
\begin{eqnarray}
D_{\mu\lambda}(k) {D^{\lambda}}_{\nu}(k)=
D_{\mu\perp}(k) {D^{\perp}}_{\nu}(k)&=& - D_{\mu\nu}(k), \nonumber \\
k^{\mu}D_{\mu\nu}(k)=0, \qquad \quad && n^{\mu}D_{\mu\nu}(k)\equiv
D_{-\nu}(k)=0, \nonumber \\ D_{\lambda\mu}(q) \,D^{\mu\nu}(k)\,
D_{\nu\rho}(q') &=& -D_{\lambda\mu}(q)D^{\mu\rho}(q').
\end{eqnarray}
The gauge field propagator $\,\,i\,D_{\mu\nu}(k)/ (k^{2}+i\epsilon)\,$ is
transverse not only to the gauge direction $n_{\mu}$ but also to
$k_{\mu}$, {\em i.e.}, it is {\it doubly-transverse}.  This leads to
appreciable simplifications in the computations in QCD.
For example, the coupling of gluons to propagators carrying high momenta is
automatic.  The absence of
collinear divergences in irreducible
diagrams in the light-cone gauge
greatly simplifies the leading-twist
factorization of soft
and hard gluonic corrections in high momentum transfer inclusive and
exclusive reactions \cite{Lepage:1980fj} since the numerators associated
with the gluon coupling only have transverse components.
The renormalization
factors in the light-cone gauge are independent of the reference direction
$n^\mu$.  Since the gluon only has physical polarization, its
renormalization factors satisfy
$Z_1=Z_3$.  Because of its explicit unitarity in each graph, the
doubly-transverse gauge is well suited for calculations identifying the
 ``pinch" effective charge \cite{Cornwall:1989gv,Brodsky:2000cr}.

The running coupling
constant and QCD
$\beta$ function have also been computed at one loop in the
doubly-transverse light-cone gauge \cite{Srivastava:2000cf}.
It is also possible to
effectively quantize QCD using light-front methods in
covariant Feynman
gauge \cite{Srivastava:2000gi}.

A
remarkable advantage of light-front quantization is that the vacuum state
$\ket{0}$ of the full QCD Hamiltonian evidently coincides with the free
vacuum.  The light-front vacuum is effectively trivial if the
interaction Hamiltonian applied to the perturbative vacuum is zero.
Note that
all particles in the Hilbert space have positive energy $k^0 =
{1\over 2}(k^+ + k^-)$, and thus positive light-front
$k^\pm$.  Since the plus momenta $\sum k^+_i$ is conserved by the
interactions, the perturbative vacuum can only couple to states with
particles in which all $k^+_i$ = 0; \ie, so called zero-mode states.  In
the case of QED, a massive electron cannot have $k^+ = 0$ unless it
also has infinite energy.  In a remarkable calculation,
Bassetto and collaborators \cite{Bassetto:1999tm} have shown that the
computation of the spectrum of $QCD(1+1)$ in equal time quantization
requires constructing the full spectrum of non perturbative contributions
(instantons).  In contrast, in the light-front quantization of gauge
theory, where the $k^+ = 0 $ singularity of the instantaneous interaction
is defined by a simple infrared regularization, one obtains
the correct spectrum of $QCD(1+1)$ without any need for vacuum-related
contributions.

In the case of QCD(3+1), the
momentum-independent four-gluon non-Abelian interaction in principle
can couple the perturbative vacuum to a state with four collinear gluons
in which all of the gluons have all components
$k^\mu_i = 0,$ thus hinting at role for zero modes in theories with
massless quanta.  In fact, zero modes of auxiliary fields are
necessary to distinguish the theta-vacua of massless
QED(1+1) \cite{Yamawaki:1998cy,McCartor:2000yy,Srivastava:1999et},
or to represent a theory in the presence of
static external boundary conditions or other constraints.  Zero-modes
provide the light-front representation of spontaneous symmetry breaking in
scalar theories \cite{Pinsky:1994yi}.

There are a number of other simplifications of the light-front
formalism:

1.  The light-front wavefunctions describe quanta which have positive
energy, positive norm, and physical polarization.  The formalism is thus
physical, and unitary.  No ghosts fields appear explicitly, even in
non-Abelian theory.  The wavefunctions are only functions of three rather
than four physical momentum variables: the light-front momentum fractions
$x_i$ and transverse momenta $k_\perp$.  The quarks and gluons each have
two physical polarization states.

2.  The set of light-front wavefunctions provide a
frame-independent, quantum-mechanical description of hadrons at the
amplitude level capable of encoding multi-quark and gluon
momentum, helicity, and flavor correlations in the form of universal
process-independent hadron wavefunctions.
Matrix elements of spacelike currents such as the spacelike
electromagnetic form factors have an exact representation in terms of
simple overlaps of the light-front wavefunctions in momentum space with
the same
$x_i$ and unchanged parton
number \cite{Drell:1970km,West:1970av,Brodsky:1980zm}. In the case of
timelike decays, such as those determined by semileptonic $B$ decay, one
needs to include contributions in which the parton number
$\Delta n =2.$ \cite{Brodsky:1999hn}.  The leading-twist off-forward parton
distributions measured in deeply virtual Compton scattering have a
similar light-front wavefunction
representation \cite{Brodsky:2000xy,Diehl:2000xz}.

3.  The high $x \to 1$ and high $k_\perp$ limits of the
hadron wavefunctions control processes and reactions in which
the hadron wavefunctions are highly stressed.  Such configurations
involve far-off-shell intermediate states and can be systematically
treated in perturbation theory \cite{Brodsky:1995kg,Lepage:1980fj}.

4.  The leading-twist structure functions $q_i(x,Q)$ and $g(x,Q)$ measured
in deep inelastic scattering can be computed from the absolute squares of
the light-front wavefunctions, integrated over the transverse momentum up
to the resolution scale $Q$.  All helicity distributions are thus
encoded in terms of the light-front wavefunctions.  The DGLAP evolution of
the structure functions can be derived from the high $k_\perp$ properties
of the light-front wavefunctions.
Thus given the light-front wavefunctions, one can compute \cite{Lepage:1980fj}
all of the leading twist helicity and
transversity distributions measured in polarized deep inelastic
lepton scattering.  For example,
the helicity-specific quark distributions at resolution $\Lambda$
correspond to
\begin{eqnarray}
q_{\lambda_q/\Lambda_p}(x, \Lambda)
&=& \sum_{n,q_a}
\int\prod^n_{j=1} {dx_j d^2 k_{\perp j}\over 16 \pi^3} \sum_{\lambda_i}
\vert \psi^{(\Lambda)}_{n/H}(x_i,\vec k_{\perp i},\lambda_i)\vert^2
\\
&& \times 16 \pi^3 \delta\left(1- \sum^n_i x_i\right) \delta^{(2)}
\left(\sum^n_i \vec k_{\perp i}\right)
\delta(x - x_q) \delta_{\lambda, \lambda_q}
\Theta(\Lambda^2 - {\cal M}^2_n)\ , \nonumber
\end{eqnarray}
where the sum is over all quarks $q_a$ which match the quantum
numbers, light-front momentum fraction $x,$ and helicity of the struck
quark.  Similarly, the transversity distributions and
off-diagonal helicity convolutions are defined as a density matrix of the
light-front wavefunctions.  This defines the LC
factorization scheme \cite{Lepage:1980fj} where the
invariant mass squared ${\cal M}^2_n = \sum_{i = 1}^n {(k_{\perp i}^2 +
m_i^2 )/ x_i}$ of the $n$ partons of the light-front wavefunctions is
limited to $ {\cal M}^2_n < \Lambda^2$

5.  The distribution of spectator
particles in the final state in the proton
fragmentation region in deep inelastic scattering at an electron-proton
collider are encoded in the light-front wavefunctions of the target
proton.  Conversely, the light-front wavefunctions can be used to describe
the coalescence of comoving quarks into final state hadrons.

6.  The light-front wavefunctions also specify the multi-quark and gluon
correlations of the hadron.  Despite the many sources
of power-law corrections to the deep inelastic cross section, certain
types of dynamical contributions will stand out at large $x_{bj}$ since
they arise from compact, highly-correlated fluctuations of the proton
wavefunction.  In particular, there are
particularly interesting dynamical ${\cal O}(1/Q^2)$ corrections which
are due to the {\it interference} of quark currents; {\it i.e.},
contributions which involve
leptons scattering amplitudes from two
different quarks of the target nucleon \cite{Brodsky:2000zu}.

7.  The higher Fock states of the light hadrons
describe the sea quark structure of the deep inelastic structure
functions, including ``intrinsic" strange\-ness and charm fluctuations
specific to the hadron's structure rather than
gluon substructure \cite{Brodsky:1980pb,Harris:1996jx}. Ladder relations
connecting state of different particle number follow from the QCD equation
of motion and lead to Regge behavior of the quark and gluon distributions at
$x \to 0$ \cite{Antonuccio:1997tw}.

8.  The gauge- and
process-independent meson and baryon valence-quark distribution amplitudes
$\phi_M(x,Q)$, and
$\phi_B(x_i,Q)$ which
control exclusive processes involving a hard scale
$Q$, including heavy quark decays, are given by the valence light-front
Fock state wavefunctions integrated over the transverse momentum up to
the resolution scale $Q$.  The evolution equations for distribution
amplitudes
follow from the perturbative high transverse momentum behavior of the
light-front wavefunctions \cite{Brodsky:1989pv}.

9.  The line-integrals
needed to defining distribution
amplitudes and structure functions as gauge invariant matrix elements of
operator products vanish in light-front gauge.

10.  Proofs of factorization theorems in hard exclusive and inclusive
reactions are greatly simplified since the propagating gluons in
light-cone gauge couple only to transverse currents; collinear
divergences are thus automatically suppressed.

11.  At high energies each light-front Fock state interacts
distinctly; \eg, Fock states with small particle number and small impact
separation have small color dipole moments and can traverse a nucleus with
minimal interactions.  This is the basis for the predictions for ``color
transparency" in hard
quasi-exclusive \cite{Brodsky:1988xz,Frankfurt:1988nt} and diffractive
reactions \cite{Bertsch:1981py,Frankfurt:1993it,Frankfurt:2000tq}.

12.  The Fock state wavefunctions of hadron can be resolved by a high
energy diffractive interaction, producing forward jets with momenta which
follow the light-front momenta of the
wavefunction \cite{Bertsch:1981py,Frankfurt:1993it,Frankfurt:2000tq}.

13.
The deuteron form factor at high $Q^2$ is sensitive to wavefunction
configurations where all six quarks overlap within an impact
separation $b_{\perp i} < {\cal O} (1/Q).$ The leading power-law
fall off predicted by QCD is $F_d(Q^2) = f(\alpha_s(Q^2))/(Q^2)^5$,
where, asymptotically, $f(\alpha_s(Q^2)) \propto
\alpha_s(Q^2)^{5+2\gamma}$ \cite{Brodsky:1976rz,Brodsky:1983vf}.
In general, the six-quark wavefunction of a deuteron
is a mixture of five different color-singlet states.  The dominant
color configuration at large distances corresponds to the usual
proton-neutron bound state.  However at small impact space
separation, all five Fock color-singlet components eventually evolve to a
state with equal weight, \ie, the deuteron wavefunction evolves to
80\%\ ``hidden color'' \cite{Brodsky:1983vf}.
The relatively large normalization of the
deuteron form factor observed at large $Q^2$ hints at sizable
hidden-color contributions \cite{Farrar:1991qi}. Hidden color components
can also play a predominant role in the reaction $\gamma d \to J/\psi p n$
at threshold if it is dominated by the multi-fusion process $\gamma g g
\to J/\psi$ \cite{Brodsky:2000zc}.  Hard exclusive nuclear processes can
also be analyzed in terms of ``reduced amplitudes" which remove the effects
of nucleon substructure.

Light-front wavefunctions are thus the frame-independent interpolating
functions between hadron and quark and gluon degrees of freedom.  Hadron
amplitudes are computed from the convolution of the light-front
wavefunctions with irreducible quark-gluon amplitudes.  More generally,
all multi-quark and gluon correlations in the bound state are represented
by the light-front wavefunctions.
The light-front Fock representation is thus a representation of the
underlying quantum field theory.  I will discuss progress in computing
light-front wavefunctions directly from QCD in Sections 9 and 10.

\section{Other Theoretical Tools}

In addition to the light-front Fock expansion, a number of other
useful theoretical tools are available to eliminate theoretical
ambiguities in QCD predictions:

(1) Conformal symmetry provides a template for
QCD predictions \cite{Brodsky:1999gm}, leading to relations between
observables which are present even in a theory which is not scale
invariant.  For example, the natural representation of distribution
amplitudes is in terms of an expansion of orthonormal conformal functions
multiplied by anomalous dimensions determined by QCD evolution
equations \cite{Brodsky:1980ny,Muller:1994hg,Braun:1999te}. Thus an
important guide in QCD analyses is to identify the underlying conformal
relations of QCD which are manifest if we drop quark masses and effects
due to the running of the QCD couplings.  In fact, if QCD has an infrared
fixed point (vanishing of the Gell Mann-Low function at low momenta), the
theory will closely resemble a scale-free conformally symmetric theory in
many applications.

(2) Commensurate scale relations \cite{Brodsky:1995eh,Brodsky:1998ua} are
perturbative QCD predictions which relate observable to observable at
fixed relative scale, such as the
``generalized Crewther relation" \cite{Brodsky:1996tb}, which connects the
Bjorken and Gross-Llewellyn
Smith deep inelastic scattering sum rules to measurements of the $e^+
e^-$ annihilation cross section.  Such relations have no renormalization
scale or scheme ambiguity.  The coefficients in the perturbative series
for commensurate scale relations are identical to those of conformal QCD;
thus no infrared renormalons are present \cite{Brodsky:1999gm}. One can
identify the required conformal coefficients at any finite order by
expanding the coefficients of the usual PQCD expansion around a formal
infrared fixed point, as in the Banks-Zak method \cite{Brodsky:2000cr}.
All non-conformal effects are absorbed by fixing the ratio of the
respective momentum transfer and energy scales.  In the case of
fixed-point theories, commensurate scale relations relate both the ratio
of couplings and the ratio of scales as the fixed point is approached
 \cite{Brodsky:1999gm}.

(3) $\alpha_V$ and Skeleton Schemes.  A physically natural scheme for
defining the QCD coupling in exclusive and other processes is the
$\alpha_V(Q^2)$ scheme defined from the potential of static heavy
quarks.  Heavy-quark lattice gauge theory can provide highly precise
values for the coupling.  All vacuum polarization corrections due to
fermion pairs are then automatically and analytically incorporated into
the Gell Mann-Low function, thus avoiding the problem of explicitly
computing and resumming quark mass corrections related to the running of
the coupling \cite{Brodsky:1998mf}. The use of a finite effective charge such
as
$\alpha_V$ as the expansion parameter also provides a basis for
regulating the infrared nonperturbative domain of the QCD coupling.  A
similar coupling and scheme can be based on an assumed skeleton expansion
of the theory \cite{Cornwall:1989gv,Brodsky:2000cr}.

(4) The Abelian Correspondence Principle.  One can consider QCD
predictions as analytic functions of the number of colors $N_C$ and
flavors $N_F$.  In particular, one can show at all orders of
perturbation theory that PQCD predictions reduce to those of an Abelian
theory at $N_C \to 0$ with ${\widehat \alpha} = C_F \alpha_s$ and
${\widehat N_F} = 2 N_F/C_F$ held fixed \cite{Brodsky:1997jk}. There is
thus a deep connection between QCD processes and their corresponding QED
analogs.

\section{Applications of light-front Wavefunctions to
Current Matrix Elements}

The light-front Fock representation
of current matrix elements has a number of simplifying properties.
The space-like local operators for the coupling of photons, gravitons and
the deep inelastic structure functions can all be expressed as overlaps
of light-front wavefunctions with the same number of Fock constituents.
This is possible since one can choose the special frame
$q^+ = 0$ \cite{Drell:1970km,West:1970av}
for space-like momentum transfer and
take matrix elements of ``plus" components of currents such as $J^+$ and
$T^{++}$.  No contributions to the current matrix elements from
vacuum fluctuations occur.  Similarly, given the
local operators for the energy-momentum tensor
$T^{\mu \nu}(x)$ and the angular momentum tensor
$M^{\mu \nu \lambda}(x)$, one can directly compute
momentum fractions, spin properties, and the form factors $A(q^2)$ and
$B(q^2)$ appearing in the coupling of gravitons
to composite systems \cite{Brodsky:2001ii}.

In the case of a spin-${1\over 2}$ composite system, the Dirac and
Pauli form factors $F_1(q^2)$ and $F_2(q^2)$ are defined by
\begin{equation}
    \langle P'| J^\mu (0) |P\rangle
       = \bar u(P')\, \Big[\, F_1(q^2)\gamma^\mu +
F_2(q^2){i\over 2M}\sigma^{\mu\alpha}q_\alpha\, \Big] \, u(P)\ ,
\label{Drell1}
\end{equation}
where $q^\mu = (P' -P)^\mu$ and $u(P)$ is the bound state spinor.
In the light-front formalism it is convenient to identify the Dirac and
Pauli form factors from the
helicity-conserving and helicity-flip vector current matrix elements of
the $J^+$ current \cite{Brodsky:1980zm}:
\begin{equation}
\VEV{P+q,\uparrow\left|\frac{J^+(0)}{2P^+}
\right|P,\uparrow} =F_1(q^2) \ ,
\label{BD1}
\end{equation}
\begin{equation}
\VEV{P+q,\uparrow\left|\frac{J^+(0)}{2P^+}\right|P,\downarrow}
=-(q^1-{\mathrm i} q^2){F_2(q^2)\over 2M}\ .
\label{BD2}
\end{equation}
The magnetic moment of a composite system is one of its
most basic properties.  The magnetic moment is defined at the $q^2 \to 0$
limit,
\begin{equation}
\mu=\frac{e}{2 M}\left[ F_1(0)+F_2(0) \right] ,
\label{DPmu}
\end{equation}
where $e$ is the charge and $M$ is the mass of the composite
system.  We use the standard light-front frame
($q^{\pm}=q^0\pm q^3$):
\begin{eqnarray}
q &=& (q^+,q^-,{\vec q}_{\perp}) = \left(0, \frac{-q^2}{P^+},
{\vec q}_{\perp}\right), \nonumber \\
P &=& (P^+,P^-,{\vec P}_{\perp}) = \left(P^+, \frac{M^2}{P^+},
{\vec 0}_{\perp}\right),
\label{LCF}
\end{eqnarray}
where $q^2=-2 P \cdot q= -{\vec q}_{\perp}^2$ is 4-momentum square
transferred by the photon.

The Pauli form factor and the anomalous magnetic moment $\kappa =
{e\over 2 M} F_2(0)$ can then be calculated from the
expression
\begin{equation}
-(q^1-{\mathrm i} q^2){F_2(q^2)\over 2M} =
\sum_a  \int
{{\mathrm d}^2 {\vec k}_{\perp} {\mathrm d} x \over 16 \pi^3}
\sum_j e_j \ \psi^{\uparrow *}_{a}(x_i,{\vec k}^\prime_{\perp
i},\lambda_i) \,
\psi^\downarrow_{a} (x_i, {\vec k}_{\perp i},\lambda_i)
{}\ ,
\label{LCmu}
\end{equation}
where the summation is over all contributing Fock states $a$ and struck
constituent charges $e_j$.  The arguments of the final-state
light-front wavefunction are
\begin{equation}
{\vec k}'_{\perp i}={\vec k}_{\perp i}+(1-x_i){\vec
q}_{\perp}
\label{kprime1}
\end{equation}
for the struck constituent and
\begin{equation}
{\vec k}'_{\perp i}={\vec k}_{\perp i}-x_i{\vec q}_{\perp}
\label{kprime2}
\end{equation}
for each spectator.
Notice that the magnetic moment must be calculated from the
spin-flip non-forward matrix element of the current.
In the ultra-relativistic limit where the radius of the system is small
compared to its Compton scale $1/M$, the anomalous magnetic moment must
vanish \cite{Brodsky:1980zm}. The light-front formalism is consistent with
this theorem.

The form factors of the energy-momentum tensor for a spin-$\half$ \
composite are defined by
\begin{eqnarray}
      \langle P'| T^{\mu\nu} (0)|P \rangle
       &=& \bar u(P')\, \Big[\, A(q^2)
       \gamma^{(\mu} \bar P^{\nu)} +
   B(q^2){i\over 2M} \bar P^{(\mu} \sigma^{\nu)\alpha}
q_\alpha \nonumber \\
   &&\qquad\qquad +  C(q^2){1\over M}(q^\mu q^\nu - g^{\mu\nu}q^2)
    \, \Big]\, u(P) \ ,
\label{Ji12}
\end{eqnarray}
where $q^\mu = (P'-P)^\mu$,
$\bar P^\mu={1\over 2}(P'+P)^\mu$,
$a^{(\mu}b^{\nu)}={1\over 2}(a^\mu b^\nu +a^\nu b^\mu)$,
and $u(P)$ is the spinor of the system.
One can also readily obtain the light-front representation
of the $A(q^2)$ and $B(q^2)$ form factors of the energy-tensor
Eq. (\ref{Ji12}) \cite{Brodsky:2001ii}.
In the interaction picture, only the non-interacting
parts of the energy momentum tensor $T^{+ +}(0)$ need to be computed:
\begin{equation}
\VEV{P+q,\uparrow\left|\frac{T^{++}(0)}{2(P^+)^2}
\right|P,\uparrow} =A(q^2)\ ,
\label{eBD1}
\end{equation}
\begin{equation}
\VEV{P+q,\uparrow\left|\frac{T^{++}(0)}{2(P^+)^2}\right|P,\downarrow}=
-(q^1-{\mathrm i} q^2){B(q^2)\over 2M}\ .
\label{eBD2}
\end{equation}
The $A(q^2)$ and $B(q^2)$ form factors
Eqs. (\ref{eBD1}) and (\ref{eBD2})
are similar to the $F_1(q^2)$ and $F_2(q^2)$ form
factors Eqs.  (\ref{BD1}) and (\ref{BD2}) with an additional
factor of the light-front momentum fraction $x=k^+/P^+$ of the struck
constituent in the integrand.  The $B(q^2)$ form factor is obtained from
the non-forward spin-flip amplitude.  The value of $B(0)$ is obtained in
the $q^2 \to 0$ limit.
The angular
momentum projection of a state is given by
\begin{equation}
\VEV{J^i} = {1\over 2} \epsilon^{i j k} \int d^3x \VEV{T^{0 k}x^j - T^{0 j} x^k}
= A(0) \VEV{L^i} + \left[A(0) + B(0)\right] \bar u(P){1\over
2}\sigma^i u(P)
\ .
\label{Ji13a}
\end{equation}
This result is derived using a wave-packet description of the state.  The
$\VEV{L^i}$ term is the orbital angular momentum of the center of mass motion
with respect to an arbitrary origin and can be dropped.  The coefficient
of the $\VEV{L^i}$ term must be 1; $A(0) = 1 $ also follows when we evaluate
the four-momentum expectation value $\VEV{P^\mu}$.
Thus the total intrinsic angular momentum
$J^z$ of a nucleon can be identified with the values of the form factors
$A(q^2)$ and
$B(q^2)$ at
$q^2= 0:$
\begin{equation}
      \VEV{J^z} = \VEV{{1\over 2} \sigma^z} \left[A(0) + B(0)\right] \ .
\label{Ji13}
\end{equation}

The
anomalous moment coupling $B(0)$ to a graviton is shown to vanish for
any composite system.  This remarkable
result, first derived by
Okun and Kobzarev \cite{Okun,Ji:1996kb,Ji:1997ek,Ji:1997nm,Teryaev:1999su},
is shown to follow directly from the Lorentz
boost properties of the light-front Fock
representation \cite{Brodsky:2001ii}.

Dae Sung Hwang, Bo-Qiang Ma, Ivan Schmidt, and
I \cite{Brodsky:2001ii} have recently shown that the light-front
wavefunctions generated by the radiative corrections to the electron in
QED provides a simple system for understanding the spin and angular
momentum decomposition of relativistic systems.  This
perturbative model also illustrates the interconnections between Fock
states of different number.  The model is patterned after the quantum
structure which occurs in the one-loop Schwinger
${\alpha / 2 \pi} $ correction to the electron magnetic
moment \cite{Brodsky:1980zm}. In effect, we can represent a spin-$\half$ ~
system as a composite of a spin-$\half$ ~ fermion and spin-one vector
boson with arbitrary masses.  A similar model has been used to
illustrate the matrix elements and evolution of light-front helicity and
orbital angular momentum operators \cite{Harindranath:1999ve}. This
representation of a composite system is particularly useful because it
is based on two constituents but yet is totally relativistic.  We can
then explicitly compute the form factors
$F_1(q^2)$ and $F_2(q^2)$ of the electromagnetic current, and the
various contributions to the form factors
$A(q^2)$ and $B(q^2)$ of the energy-momentum tensor.

\vspace{.5cm}
\begin{figure}[htb]
\begin{center}
\leavevmode
\epsfbox{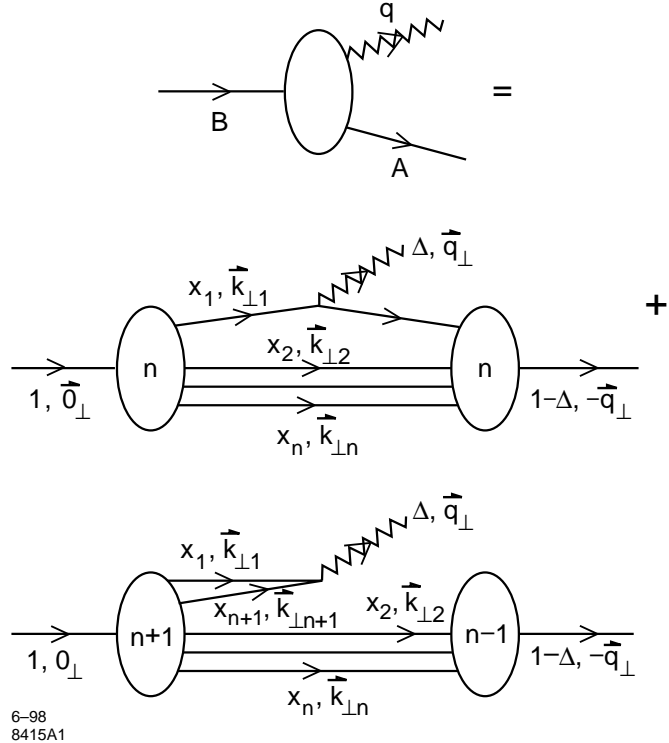}
\end{center}
\caption[*]{Exact representation of electroweak decays and time-like form
factors in the
light-front Fock representation.
}
\label{fig1}
\end{figure}

Another remarkable advantage of the light-front formalism is that
exclusive semileptonic
$B$-decay amplitudes such as $B\rightarrow A \ell \bar{\nu}$ can also be
evaluated exactly \cite{Brodsky:1999hn}.
The time-like decay matrix elements require the computation of the
diagonal matrix element $n \rightarrow n$ where parton number is
conserved, and the off-diagonal $n+1\rightarrow n-1$ convolution where
the current operator annihilates a $q{\bar{q'}}$ pair in the initial $B$
wavefunction.  See Fig.  \ref{fig1}.  This term is a consequence of the
fact that the time-like decay $q^2 = (p_\ell + p_{\bar{\nu}} )^2 > 0$
requires a positive light-front momentum fraction $q^+ > 0$.  Conversely
for space-like currents, one can choose $q^+=0$, as in the
Drell-Yan-West representation of the space-like electromagnetic form
factors.  However, as can be seen from the explicit analysis of the form
factor in a perturbative model, the off-diagonal convolution can yield a
nonzero $q^+/q^+$ limiting form as $q^+ \rightarrow 0$.  This extra term
appears specifically in the case of ``bad" currents such as $J^-$ in
which the coupling to $q\bar q$ fluctuations in the light-front
wavefunctions are favored.  In effect, the $q^+ \rightarrow 0$ limit
generates $\delta(x)$ contributions as residues of the $n+1\rightarrow
n-1$ contributions.  The necessity for such ``zero mode" $\delta(x)$ terms
has been noted by Chang, Root and Yan \cite{Chang:1973xt},
Burkardt \cite{Burkardt:1989wy}, and Ji and Choi \cite{Choi:1998nf}.

The off-diagonal $n+1 \rightarrow n-1$ contributions give a new
perspective for the physics of $B$-decays.  A semileptonic decay
involves not only matrix elements where a quark changes flavor, but also
a contribution where the leptonic pair is created from the annihilation
of a $q {\bar{q'}}$ pair within the Fock states of the initial $B$
wavefunction.  The semileptonic decay thus can occur from the
annihilation of a nonvalence quark-antiquark pair in the initial hadron.
This feature will carry over to exclusive hadronic $B$-decays, such as
$B^0 \rightarrow \pi^-D^+$.  In this case the pion can be produced from
the coalescence of a $d\bar u$ pair emerging from the initial higher
particle number Fock wavefunction of the $B$.  The $D$ meson is then
formed from the remaining quarks after the internal exchange of a $W$
boson.

In principle, a precise evaluation of the hadronic matrix elements
needed for $B$-decays and other exclusive electroweak decay amplitudes
requires knowledge of all of the light-front Fock wavefunctions of the
initial and final state hadrons.  In the case of model gauge theories
such as QCD(1+1) \cite{Hornbostel:1990fb} or
collinear QCD \cite{Antonuccio:1995fs} in one-space and one-time dimensions,
the complete
evaluation of the light-front wavefunction is possible for each baryon or
meson bound-state using the DLCQ method.  It would be interesting to use
such solutions as a model for physical $B$-decays.

\section{Light-front Representation of Deeply Virtual
Compton Scattering}

The virtual Compton scattering process ${d\sigma\over dt}(\gamma^*
p \to \gamma p)$ for large initial photon virtuality
$q^2=-Q^2$  has extraordinary sensitivity to
fundamental features of the proton's structure.  Even though the final
state photon is on-shell, the deeply virtual process probes the
elementary quark structure of the proton near the light cone as an
effective local current.  In contrast to deep inelastic scattering, which
measures only the absorptive part of the forward virtual Compton
amplitude $Im {\cal T}_{\gamma^* p \to
\gamma^* p}$, deeply virtual Compton scattering allows the measurement of
the phase and spin structure of proton matrix elements for general
momentum transfer squared $t$.  In addition, the interference of the
virtual Compton amplitude and Bethe-Heitler
wide angle scattering Bremsstrahlung amplitude where the photon is
emitted from the lepton line leads to an electron-positron asymmetry in
the ${e^\pm  p \to e^\pm \gamma p}$ cross section which is proportional
to the real part of the Compton
amplitude \cite{Brodsky:1972zh,Brodsky:1972vv,Brodsky:1973hm}.  The deeply
virtual Compton amplitude
$\gamma^* p \to \gamma p$ is related by crossing to another important
process
$\gamma^* \gamma
\to $ hadron pairs at fixed invariant mass which can be measured
in electron-photon collisions \cite{Diehl:2000uv}.

To leading order in $1/Q$, the deeply virtual Compton scattering
amplitude $\gamma^*(q) p(P) \to \gamma(q') p(P')$ factorizes as the
convolution in $x$ of the
amplitude $t^{\mu \nu}$ for hard Compton scattering on a quark line with
the generalized Compton form factors $H(x,t,\zeta),$ $ E(x,t,\zeta)$,
$\tilde H(x,t,\zeta),$ and $\tilde E(x,t,\zeta)$ of the target
proton \cite{Ji:1996kb,Ji:1997ek,Radyushkin:1996nd,Ji:1998xh,%
Guichon:1998xv,Vanderhaeghen:1998uc,Radyushkin:1999es,%
Collins:1999be,Diehl:1999tr,Diehl:1999kh,Blumlein:2000cx,Penttinen:2000dg}.
Here
$x$ is the light-front momentum fraction of the struck quark, and
$\zeta= Q^2/2 P\cdot q$ plays the role of the Bjorken variable.  The
square of the four-momentum transfer from the proton is given by
$t=\Delta^2\ = \ 2P\cdot \Delta\  =\
-{(\zeta^2M^2+{\vec \Delta_\perp}^2)\over (1-\zeta)}\ $ ,
where $\Delta $ is the difference of initial and final momenta of the
proton ($P=P'+\Delta$).
We will be interested in deeply virtual Compton scattering where $q^2$ is
large compared to the masses and $t$.  Then, to leading order in $1/Q^2$,
${-q^2\over 2P_I\cdot q}=\zeta\ .$
Thus $\zeta$ plays the role of the Bjorken variable in deeply virtual
Compton scattering.
For a fixed value of $-t$, the allowed range of $\zeta$ is given by
\begin{equation}
0\ \le\ \zeta\ \le\
{(-t)\over 2M^2}\ \ \left( {\sqrt{1+{4M^2\over (-t)}}}\ -\ 1\ \right)\ .
\label{nn4}
\end{equation}
The form factor $H(x,t,\zeta)$ describes the proton response
when the helicity of the proton is unchanged, and
$E(x,t,\zeta)$ is for the case when the proton helicity is flipped.  Two
additional functions $\tilde H(x,t,\zeta),$ and $\tilde E(x,t,\zeta)$
appear, corresponding to the dependence of the Compton amplitude on
quark helicity.

\vspace{.5cm}
\begin{figure}[htb]
\begin{center}
\leavevmode
{\epsfxsize=4in\epsfbox{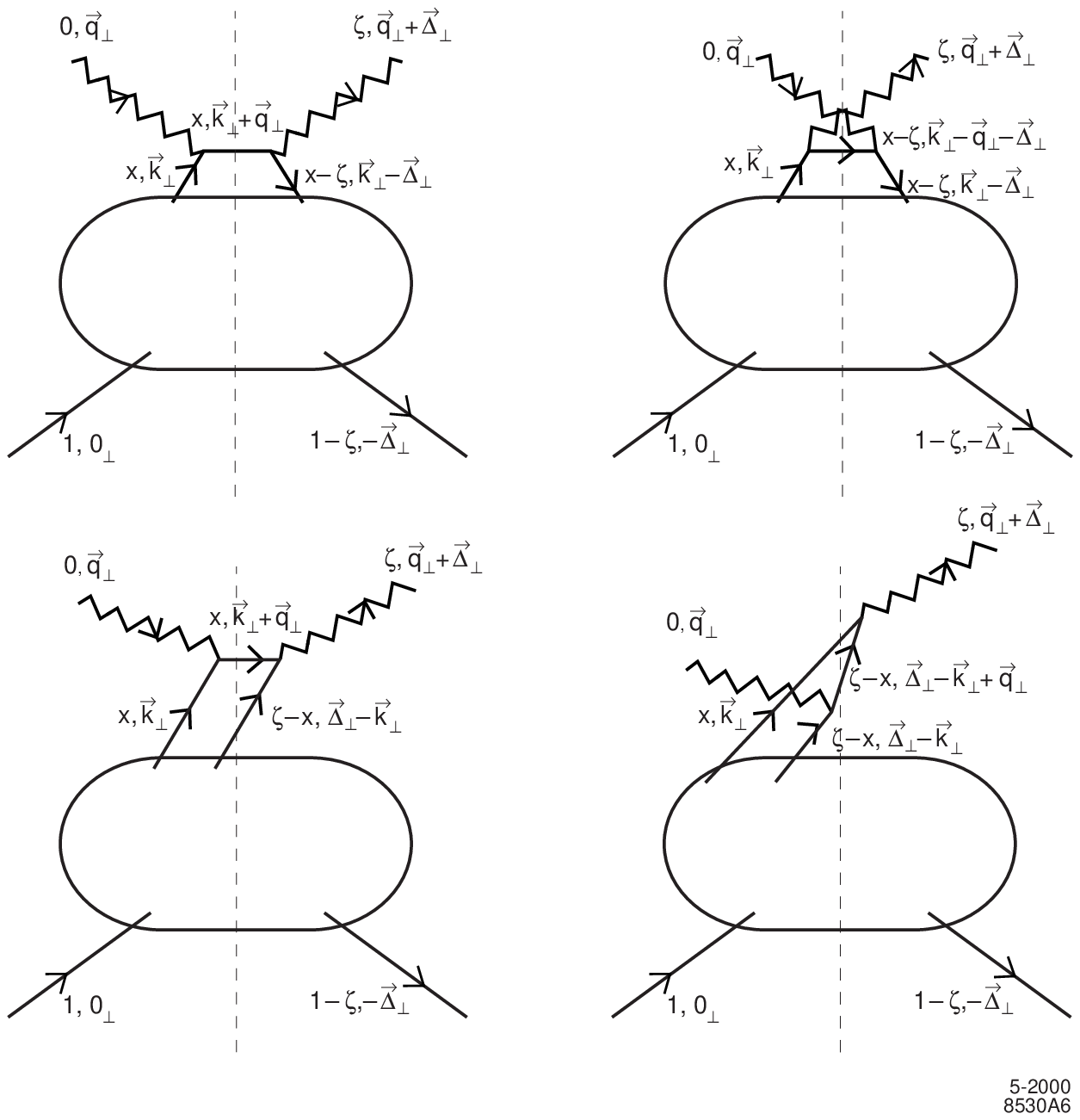}}
\end{center}
\caption[*]{Light-cone time-ordered contributions to deeply virtual
Compton scattering.  Only the contributions of leading twist in $1/q^2$
are illustrated.  These contributions illustrate the factorization
property of the leading twist amplitude.}
\label{fig:3}
\end{figure}

Recently, Markus Diehl, Dae Sung Hwang and I \cite{Brodsky:2000xy} have
shown how the deeply virtual Compton amplitude can be evaluated
explicitly in the Fock state representation using the matrix elements of
the currents and the boost properties of the light-front wavefunctions.
For the $n \to n$ diagonal term ($\Delta n = 0$), the arguments of the
final-state hadron wavefunction are
$x_1-\zeta \over 1-\zeta$,
${\vec{k}}_{\perp 1} - {1-x_1\over 1-\zeta} {\vec{\Delta}}_\perp$ for
the struck quark
and $x_i\over 1-\zeta$,
${\vec{k}}_{\perp i} + {x_i\over 1-\zeta} {\vec{\Delta}}_\perp$
for the $n-1$ spectators.
We thus obtain formulae for the diagonal (parton-number-conserving)
contribution to the generalized form factors for deeply virtual Compton
amplitude in the domain \cite{Diehl:1999kh,Diehl:1999tr,Muller:1994fv}
$\zeta\le x_1\le 1$:\\
\begin{eqnarray}
&&{\sqrt{1-\zeta}}f_{1\, (n\to n)}(x_1,t,\zeta)\,
-\, {\zeta^2\over 4{\sqrt{1-\zeta}}} f_{2\, (n\to n)}(x_1,t,\zeta)
\nonumber\\
 &=&
\sum_{n, ~ \lambda}
\prod_{i=1}^{n}
\int^1_0 dx_{i(i\ne 1)} \int {d^2{\vec{k}}_{\perp i} \over 2 (2\pi)^3 }
~ \delta\left(1-\sum_{j=1}^n x_j\right) ~ \delta^{(2)}
\left(\sum_{j=1}^n {\vec{k}}_{\perp j}\right)  \nonumber\\[1ex]
&&\times
\psi^{\uparrow \ *}_{(n)}(x^\prime_i, {\vec{k}}^\prime_{\perp i},\lambda_i)
~ \psi^{\uparrow}_{(n)}(x_i, {\vec{k}}_{\perp i},\lambda_i)
(\sqrt{1-\zeta})^{1-n},
\label{t1}
\end{eqnarray}
\begin{eqnarray}
&&
{\sqrt{1-\zeta}}\,\left(\, 1+{\zeta\over 2(1-\zeta)}\,\right)\,
{(\Delta^1-{\mathrm i} \Delta^2)\over 2M}f_{2\, (n\to n)}(x_1,t,\zeta)
\nonumber\\
 &=&
\sum_{n, ~ \lambda}
\prod_{i=1}^{n}
\int^1_0 dx_{i(i\ne 1)} \int {d^2{\vec{k}}_{\perp i} \over 2 (2\pi)^3 }
~ \delta\left(1-\sum_{j=1}^n x_j\right) ~ \delta^{(2)}
\left(\sum_{j=1}^n {\vec{k}}_{\perp j}\right)  \nonumber\\[1ex]
&&\qquad\qquad\qquad\times
\psi^{\uparrow \ *}_{(n)}(x^\prime_i, {\vec{k}}^\prime_{\perp i},\lambda_i)
~ \psi^{\downarrow}_{(n)}(x_i, {\vec{k}}_{\perp i},\lambda_i)
(\sqrt{1-\zeta})^{1-n} ,
\label{t1f2}
\end{eqnarray}
where
\begin{equation}
\left\{ \begin{array}{lll}
x^\prime_1 = {x_1-\zeta \over 1-\zeta}\, ,\
&{\vec{k}}^\prime_{\perp 1} ={\vec{k}}_{\perp 1}
- {1-x_1\over 1-\zeta} {\vec{\Delta}}_\perp
&\mbox{for the struck quark,}\\[1ex]
x^\prime_i = {x_i\over 1-\zeta}\, ,\
&{\vec{k}}^\prime_{\perp i} ={\vec{k}}_{\perp i}
+ {x_i\over 1-\zeta} {\vec{\Delta}}_\perp
&\mbox{for the $ (n-1)$ spectators.}
\end{array}\right.
\label{t2}
\end{equation}
A sum over all possible helicities $\lambda_i$ is understood.
If quark masses are neglected, the currents conserve
helicity.
We also can check that $\sum_{i=1}^n x^\prime_i = 1$,
$\sum_{i=1}^n {\vec{k}}^\prime_{\perp i} = {\vec{0}}_\perp$.

For the $n+1 \to n-1$ off-diagonal term ($\Delta n = -2$),
consider the case where
partons $1$ and
$n+1$ of the initial wavefunction annihilate into the current leaving
$n-1$ spectators.
Then $x_{n+1} = \zeta - x_{1}$,
${\vec{k}}_{\perp n+1} = {\vec{\Delta}}_\perp-{\vec{k}}_{\perp 1}$.
The remaining $n-1$ partons have total momentum
$((1-\zeta)P^+, -{\vec{\Delta}}_{\perp})$.
The final wavefunction then has arguments
$x^\prime_i = {x_i \over 1- \zeta}$ and
${\vec{k}}^\prime_{\perp i}=
{\vec{k}}_{\perp i} + {x_i\over 1-\zeta} {\vec{\Delta}}_\perp$.
We thus obtain the formulae for the off-diagonal matrix element
of the Compton amplitude in the domain $0\le x_1\le \zeta$:\\
\begin{eqnarray}
&&{\sqrt{1-\zeta}}f_{1\, (n+1\to n-1)}(x_1,t,\zeta)\,
-\, {\zeta^2\over 4{\sqrt{1-\zeta}}} f_{2\, (n+1\to n-1)}(x_1,t,\zeta)
\nonumber\\
 &=&
\sum_{n, ~ \lambda}
\int^1_0 dx_{n+1}
\int {d^2{\vec{k}}_{\perp 1} \over 2 (2\pi)^3 }
\int {d^2{\vec{k}}_{\perp n+1} \over 2 (2\pi)^3 }
\prod_{i=2}^{n}
\int^1_0 dx_{i} \int {d^2{\vec{k}}_{\perp i} \over 2 (2\pi)^3 }
\nonumber\\[2ex]
&&\times \delta\left(1-\sum_{j=1}^{n+1} x_j\right) ~
\delta^{(2)}\left(\sum_{j=1}^{n+1} {\vec{k}}_{\perp j}\right)
[\sqrt{1-\zeta}]^{1-n}
\nonumber\\[2ex]
&&\times\psi^{\uparrow\ *}_{(n-1)}
(x^\prime_i,{\vec{k}}^\prime_{\perp i},\lambda_i)
~ \psi^{\uparrow}_{(n+1)}(\{x_1, x_i, x_{n+1} = \zeta - x_{1}\},
\nonumber\\[2ex]
&&\qquad\ \ \{ {\vec{k}}_{\perp 1},
{\vec{k}}_{\perp i},
{\vec{k}}_{\perp n+1} = {\vec{\Delta}}_\perp-{\vec{k}}_{\perp 1}\},
\{\lambda_1,\lambda_{i},\lambda_{n+1} = - \lambda_{1}\})
,
\end{eqnarray}
\begin{eqnarray}
&&
{\sqrt{1-\zeta}}\,\Big(\, 1+{\zeta\over 2(1-\zeta)}\,\Big)\,
{(\Delta^1-{\mathrm i} \Delta^2)\over 2M}f_{2\, (n+1\to n-1)}(x_1,t,\zeta)
\nonumber\\
 &=&
\sum_{n, ~ \lambda}
\int^1_0 dx_{n+1}
\int {d^2{\vec{k}}_{\perp 1} \over 2 (2\pi)^3 }
\int {d^2{\vec{k}}_{\perp n+1} \over 2 (2\pi)^3 }
\prod_{i=2}^{n}
\int^1_0 dx_{i} \int {d^2{\vec{k}}_{\perp i} \over 2 (2\pi)^3 }
\nonumber\\[2ex]
&&\qquad\qquad\qquad\times \delta\left(1-\sum_{j=1}^{n+1} x_j\right) ~
\delta^{(2)}\left(\sum_{j=1}^{n+1} {\vec{k}}_{\perp j}\right)
[\sqrt{1-\zeta}]^{1-n}
\nonumber\\[2ex]
&&\qquad\qquad\qquad\times\psi^{\uparrow\ *}_{(n-1)}
(x^\prime_i,{\vec{k}}^\prime_{\perp i},\lambda_i)
~ \psi^{\downarrow}_{(n+1)}(\{x_1, x_i, x_{n+1} = \zeta - x_{1}\},
\nonumber\\[2ex]
&&\qquad\qquad\ \ \{ {\vec{k}}_{\perp 1},
{\vec{k}}_{\perp i},
{\vec{k}}_{\perp n+1} = {\vec{\Delta}}_\perp-{\vec{k}}_{\perp 1}\},
\{\lambda_1,\lambda_{i},\lambda_{n+1} = - \lambda_{1}\})
,
\end{eqnarray}
where $i=2,3,\cdots ,n$
label the $n-1$ spectator
partons which appear in the final-state hadron wavefunction
with
\begin{equation}
x^\prime_i = {x_i\over 1-\zeta}\, ,\qquad
{\vec{k}}^\prime_{\perp i} ={\vec{k}}_{\perp i}
+ {x_i\over 1-\zeta} {\vec{\Delta}}_\perp \ .
\end{equation}
We can again check that the arguments of the final-state wavefunction
satisfy
$\sum_{i=2}^n x^\prime_i = 1$,
$\sum_{i=2}^n {\vec{k}}^\prime_{\perp i} = {\vec{0}}_\perp$.

The above representation is the general form for the generalized form
factors of the deeply virtual Compton amplitude for any composite system.
Thus given the light-front Fock state wavefunctions of the eigensolutions
of the light-front Hamiltonian, we can compute the amplitude for virtual
Compton scattering including all spin correlations.  The formulae are
accurate to leading order in
$1/Q^2$.  Radiative corrections to the quark Compton amplitude of order
$\alpha_s(Q^2)$ from diagrams in which a hard gluon interacts between
the two photons have also been neglected.

\section{Applications of QCD Factorization to Hard QCD
 Processes}

Factorization theorems for hard exclusive, semi-exclusive, and
diffractive processes allow the separation of soft
non-perturbative dynamics of the bound state hadrons from the hard
dynamics of a perturbatively-calculable quark-gluon scattering
amplitude.  The factorization of inclusive reactions is reviewed in ref.
For reviews and bibliography of exclusive process calculations in QCD
(see Ref. \cite{Brodsky:1989pv,Brodsky:2000dr}).

The light-front
formalism provides a physical factorization scheme which conveniently
separates and factorizes soft non-perturbative physics from hard
perturbative dynamics in both exclusive and
inclusive reactions \cite{Lepage:1980fj,Lepage:1979zb}.

In hard inclusive
reactions all intermediate states are divided according to $\M^2_n <
\Lambda^2 $ and
$\M^2_n >
\Lambda^2 $ domains.  The lower mass regime is associated with the quark
and gluon distributions defined from the absolute squares of the LC
wavefunctions in the light cone factorization scheme.  In the high
invariant mass regime, intrinsic transverse momenta can be ignored, so
that the structure of the process at leading power has the form of hard
scattering on collinear quark and gluon constituents, as in the parton
model.  The attachment of gluons from the LC wavefunction to a propagator
in a hard subprocess is power-law suppressed in LC gauge, so that the
minimal quark-gluon particle-number subprocesses dominate.  It is then
straightforward to derive the DGLAP equations from the evolution of the
distributions with $\log \Lambda^2$.
The
anomaly contribution to singlet helicity structure function $g_1(x,Q)$
can be explicitly identified in the LC factorization scheme as due to the
$\gamma^* g \to q
\bar q$ fusion process.  The anomaly contribution would be zero if the
gluon is on shell.  However, if the off-shellness of the state is larger
than the quark pair mass, one obtains the usual anomaly
contribution \cite{Bass:1999rn}.

In exclusive amplitudes, the LC wavefunctions are the interpolating
amplitudes connecting the quark and gluons to the hadronic
states.  In an
exclusive amplitude involving a hard scale $Q^2$ all intermediate states
can be divided according to
$\M^2_n <
\Lambda^2 < Q^2 $ and $\M^2_n < \Lambda^2 $ invariant mass domains.  The
high invariant mass contributions to the amplitude has the structure of a
hard scattering process
$T_H$ in which the hadrons are replaced by their respective (collinear)
quarks and gluons.  In light-cone gauge only the minimal Fock states
contribute to the leading power-law fall-off of the exclusive amplitude.
The wavefunctions in the lower invariant mass domain can be integrated up
to an arbitrary intermediate invariant mass cutoff $\Lambda$.  The
invariant mass domain beyond this cutoff is included in the hard
scattering amplitude
$T_H$.  The $T_H$ satisfy dimensional
counting rules \cite{Brodsky:1975vy}. Final-state and initial state
corrections from gluon attachments to lines connected to the
color-singlet distribution amplitudes cancel at leading twist.  Explicit
examples of perturbative QCD factorization will be discussed in more
detail in the next section.

The key non-perturbative input for exclusive
processes is thus the gauge and frame independent hadron distribution
amplitude \cite{Lepage:1979zb,Lepage:1980fj} defined as the integral of
the valence (lowest particle number) Fock wavefunction;
\eg\ for the pion
\begin{equation}
\phi_\pi (x_i,\Lambda) \equiv \int d^2k_\perp\, \psi^{(\Lambda)}_{q\bar
q/\pi} (x_i, \vec k_{\perp i},\lambda)
\label{eq:f1a}
\end{equation}
where the global cutoff $\Lambda$ is identified with the resolution $Q$.
The distribution amplitude controls leading-twist exclusive amplitudes
at high momentum transfer, and it can be related to the gauge-invariant
Bethe-Salpeter wavefunction at equal light-cone time.  The
logarithmic evolution of hadron distribution amplitudes
$\phi_H (x_i,Q)$ can be derived from the perturbatively-computable tail
of the valence light-front wavefunction in the high transverse momentum
regime \cite{Lepage:1979zb,Lepage:1980fj}. The conformal basis for the
evolution of the three-quark distribution amplitudes for
the baryons~ \cite{Lepage:1979za} has recently been obtained by V. Braun
\etal \cite{Braun:1999te}.

The existence of an exact formalism
provides a basis for systematic approximations and a control over neglected
terms.  For example, one can analyze exclusive semi-leptonic
$B$-decays which involve hard internal momentum transfer using a
perturbative QCD
formalism \cite
{Szczepaniak:1990dt,Szczepaniak:1996xg,%
Beneke:1999br,Keum:2000ph,Keum:2000wi,Li:2000hh}
patterned after the perturbative analysis of form
factors at large momentum transfer.  The hard-scattering
analysis again proceeds
by writing each hadronic wavefunction
as a sum of soft and hard contributions
\begin{equation}
\psi_n = \psi^{{\rm soft}}_n (\M^2_n < \Lambda^2) + \psi^{{\rm hard}}_n
(\M^2_n >\Lambda^2) ,
\end{equation}
where $\M^2_n $ is the invariant mass of the partons in the $n$-particle
Fock state and
$\Lambda$ is the separation scale.
The high internal momentum contributions to the wavefunction $\psi^{{\rm
hard}}_n $ can be calculated systematically from QCD perturbation theory
by iterating the gluon exchange kernel.  The contributions from
high momentum transfer exchange to the
$B$-decay amplitude can then be written as a
convolution of a hard-scattering
quark-gluon scattering amplitude $T_H$ with the distribution
amplitudes $\phi(x_i,\Lambda)$, the valence wavefunctions obtained by
integrating the
constituent momenta up to the separation scale
${\cal M}_n < \Lambda < Q$.  Furthermore in processes such as $B \to \pi
D$ where the pion is effectively produced as a rapidly-moving small Fock
state with a small color-dipole interactions,  final state interactions
are suppressed by color transparency.  This is the basis for the
perturbative hard-scattering
analyses \cite{Szczepaniak:1990dt,Beneke:1999br,Keum:2000ph,%
Keum:2000wi,Li:2000hh}.
In a systematic
analysis, one can identify the hard PQCD contribution as well as the soft
contribution from the convolution of the light-front wavefunctions.
Furthermore, the hard-scattering contribution can be systematically improved.

Given the solution
for the hadronic wavefunctions $\psi^{(\Lambda)}_n$ with $\M^2_n <
\Lambda^2$, one can construct the wavefunction in the hard regime with
$\M^2_n > \Lambda^2$ using projection operator techniques.  The
construction can be done perturbatively in QCD since only high invariant mass,
far off-shell matrix elements are involved.  One can use this method to
derive the physical properties of the LC wavefunctions and their matrix elements
at high invariant mass.  Since $\M^2_n = \sum^n_{i=1}
\left(\frac{k^2_\perp+m^2}{x}\right)_i $, this method also allows the derivation
of the asymptotic behavior of light-front wavefunctions at large $k_\perp$,
which
in turn leads to predictions for the fall-off of form factors and other
exclusive
matrix elements at large momentum transfer, such as the quark counting rules
for predicting the nominal power-law fall-off of two-body scattering amplitudes
at fixed
$\theta_{cm}$ \cite{Brodsky:1975vy} and
helicity selection rules \cite{Brodsky:1981kj}. The phenomenological
successes of these rules
can be understood within QCD if the coupling
$\alpha_V(Q)$ freezes in a range of relatively
small momentum transfer \cite{Brodsky:1998dh}.

\section{Two-Photon Processes}

The simplest and perhaps the most elegant illustration of an exclusive
reaction in QCD is the evaluation of the photon-to-pion transition form
factor $F_{\gamma \to \pi}(Q^2)$ \cite{Lepage:1980fj,Brodsky:1981rp} which
is measurable in single-tagged two-photon
$ee \to ee \pi^0$ reactions.
The form factor is
defined via the invariant amplitude
$
\Gamma^\mu = -ie^2 F_{\pi \gamma}(Q^2) \epsilon^{\mu \nu \rho \sigma}
p^\pi_\nu \epsilon_\rho q_\sigma \ .$
As in inclusive
reactions, one must specify a factorization scheme which divides the
integration regions of the loop integrals into hard and soft momenta,
compared to the resolution scale $\tilde Q$.
At leading twist, the transition form factor then factorizes as a
convolution of the
$\gamma^* \gamma \to q \bar q$ amplitude (where the quarks are
collinear with the final state pion) with the valence light-front
wavefunction of the pion:
\begin{equation}
F_{\gamma M}(Q^2)= {4 \over \sqrt 3}\int^1_0 dx \phi_M(x,\tilde Q)
T^H_{\gamma \to M}(x,Q^2) .
\label{transitionformfactor}
\end{equation}
The hard scattering amplitude for $\gamma\gamma^*\to q \bar q$
is
$
T^H_{\gamma M}(x,Q^2) = { [(1-x) Q^2]^{-1}}\times\break \left(1 +
{\cal O}(\alpha_s)\right).
$
The leading QCD corrections have been
computed by Braaten \cite{Braaten:1987yy}.
The evaluation of the next-to-leading corrections in the physical
$\alpha_V$ scheme is given in Ref. \cite{Brodsky:1998dh}.
For the
asymptotic distribution amplitude $\phi^{\rm asympt}_\pi (x) =
\sqrt 3 f_\pi x(1-x)$ one predicts
$
Q^2 F_{\gamma \pi}(Q^2)= 2 f_\pi \left(1 - {5\over3}
{\alpha_V(Q^*)\over \pi}\right)$ where $Q^*= e^{-3/2} Q$ is the BLM scale
for the pion form factor.  The PQCD predictions have
been tested in measurements of $e \gamma \to e \pi^0$ by the CLEO
collaboration \cite{Gronberg:1998fj}.
See Fig. \ref{Fig:DalleyCleo} (b).
The observed
flat scaling of the $Q^2 F_{\gamma \pi}(Q^2)$ data from $Q^2 = 2$
to $Q^2 = 8$ GeV$^2$ provides an important confirmation of the
applicability of leading twist QCD to this process.  The magnitude of
$Q^2 F_{\gamma \pi}(Q^2)$ is remarkably consistent with the predicted
form, assuming the asymptotic distribution amplitude and including the
LO QCD radiative correction with $\alpha_V(e^{-3/2} Q)/\pi \simeq
0.12$.  One
could allow for some broadening of the distribution amplitude with a
corresponding increase in the value of $\alpha_V$ at small scales.
Radyushkin \cite{Radyushkin:1995pj}, Ong \cite{Ong:1995gs},
and Kroll \cite{Kroll:1996jx} have also noted that the scaling and
normalization of
the photon-to-pion transition form factor tends to favor the asymptotic
form for the pion distribution amplitude and rules out broader
distributions such as the two-humped form suggested by
QCD sum rules \cite{Chernyak:1984ej}.

The two-photon annihilation process $\gamma^* \gamma \to $
hadrons, which is measurable in single-tagged $e^+ e^- \to e^+ e^- {\rm
hadrons}$ events, provides a semi-local probe of
$C=+$ hadron systems $\pi^0, \eta^0, \eta^\prime, \eta_c, \pi^+ \pi^-$,
etc.  The $\gamma^* \gamma
\to \pi^+
\pi^-$ hadron pair process is related to virtual Compton
scattering on a pion target by crossing.  The leading twist amplitude is
sensitive to the
$1/x - 1/(1-x)$ moment of the two-pion distribution amplitude coupled
to two valence quarks \cite{Muller:1994fv,Diehl:2000uv}.

\begin{figure}[htb]
\begin{center}
\leavevmode
{\epsfxsize=5.5in\epsfbox{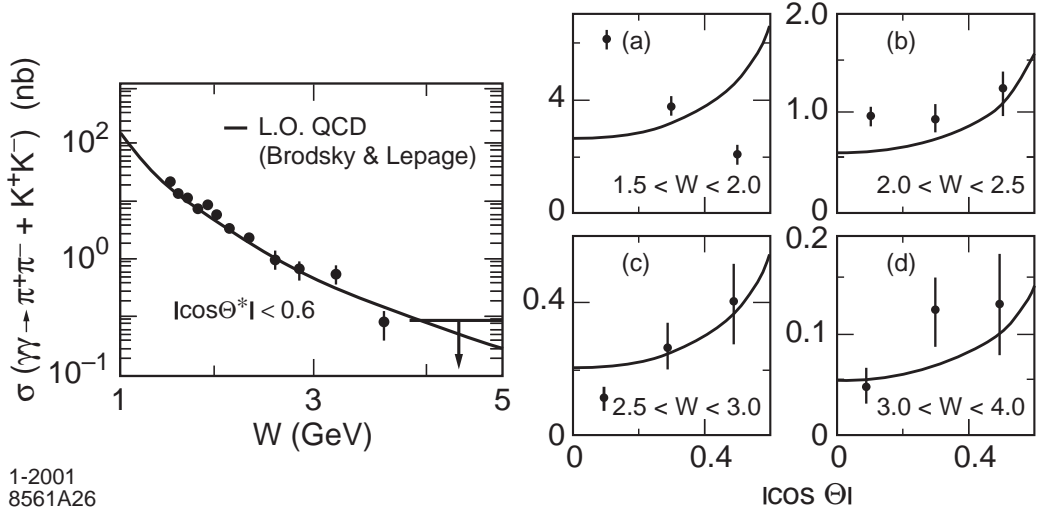}}
\end{center}
\caption[*]{Comparison of the sum of $\gamma \gamma \rightarrow \pi^+
\pi^-$ and
$\gamma \gamma \rightarrow K^+ K^-$ meson pair production cross
sections with the scaling and angular distribution of the perturbative QCD
prediction \cite{Brodsky:1981rp}.  The data are from the CLEO
collaboration \cite{Paar}.}
\label{Fig:CLEO}
\end{figure}

Two-photon reactions, $\gamma \gamma \to H \bar H$ at large s = $(k_1 +
k_2)^2$ and fixed $\theta_{\rm cm}$,
provide a particularly important laboratory for testing QCD since
these cross-channel ``Compton" processes are the simplest
calculable large-angle exclusive hadronic scattering reactions.
The helicity structure, and often even the absolute normalization can be
rigorously computed for each two-photon channel \cite{Brodsky:1981rp}.
In the case of meson pairs, dimensional counting predicts that for large
$s$, $s^4 d\sigma/dt(\gamma
\gamma \to M \bar M$ scales at fixed $t/s$ or $\theta_{\rm c.m.}$ up to
factors of $\ln s/\Lambda^2$.
The angular dependence of the $\gamma \gamma \to H \bar
H$ amplitudes can be used to determine the shape of the
process-independent distribution amplitudes, $\phi_H(x,Q)$.
An important feature of the $\gamma \gamma \to M \bar M$
amplitude for meson pairs is that the contributions of Landshoff pitch
singularities are power-law suppressed at the Born level -- even before
taking into account Sudakov form factor suppression.  There are also no
anomalous contributions from the $x \to 1$ endpoint integration region.
Thus, as in the calculation of the meson form factors, each fixed-angle
helicity amplitude can be written to leading order in $1/Q$ in the
factorized form $[Q^2 = p_T^2 = tu/s; \tilde Q_x = \min(xQ,(l-x)Q)]$:
\begin{equation}{\cal M}_{\gamma \gamma\to M \bar M}
= \int^1_0 dx \int^1_0 dy
\phi_{\bar M}(y,\tilde Q_y) T_H(x,y,s,\theta_{\rm c.m.}
\phi_{M}(x,\tilde Q_x) , \end{equation}
where $T_H$ is the hard-scattering amplitude $\gamma \gamma \to (q \bar
q) (q \bar q)$ for the production of the valence quarks collinear with
each meson, and
$\phi_M(x,\tilde Q)$ is the amplitude for
finding the valence $q$ and $\bar q$ with light-front fractions of the
meson's momentum, integrated over transverse momenta $k_\perp < \tilde
Q.$ The contribution of non-valence Fock states are power-law suppressed.
Furthermore, the helicity-selection rules \cite{Brodsky:1981kj}
of perturbative QCD predict that
vector mesons are produced with opposite helicities to leading order in
$1/Q$ and all orders in $\alpha_s$.
The dependence in $x$ and $y$ of
several terms in $T_{\lambda, \lambda'}$ is quite similar to that
appearing in the meson's electromagnetic form factor.
Thus much of the
dependence on
$\phi_M(x,Q)$ can be eliminated by expressing it in terms of the
meson form factor.
In fact, the ratio of the
$\gamma
\gamma
\to \pi^+
\pi^-$ and $e^+ e^- \to \mu^+ \mu^-$ amplitudes at large $s$ and fixed
$\theta_{CM}$ is nearly insensitive to the
running coupling and the shape of the pion distribution amplitude:
\begin{equation}{{d\sigma \over dt }(\gamma \gamma \to \pi^+ \pi^-)
\over {d\sigma \over dt }(\gamma \gamma \to \mu^+ \mu^-)}
\sim {4 \vert F_\pi(s) \vert^2 \over 1 - \cos^2 \theta_{\rm c.m.} }
.\end{equation}
The
comparison of the PQCD prediction for the sum of $\pi^+ \pi^-$ plus $K^+
K^-$ channels with recent CLEO data \cite{Paar} is shown in Fig.
\ref{Fig:CLEO}.
The CLEO data
for charged pion and kaon pairs show a clear transition to the
scaling and angular distribution predicted by PQCD \cite{Brodsky:1981rp}
for
$W = \sqrt(s_{\gamma \gamma} > 2$ GeV.  See Fig. \ref{Fig:CLEO}.  It is
clearly
important to measure the magnitude and angular dependence of the
two-photon production of neutral pions and
$\rho^+ \rho^-$ cross sections in view of the strong
sensitivity of these channels to the shape of meson distribution
amplitudes.
QCD also predicts that the production cross section for charged
$\rho$-pairs (with any helicity) is much larger that for that of
neutral $\rho$ pairs, particularly at large $\theta_{\rm c.m.}$ angles.
Similar predictions are possible for other helicity-zero mesons.
The cross sections for Compton scattering on protons and the
crossed reaction $\gamma \gamma \to p \bar p$ at high momentum transfer
have also been
evaluated \cite{Farrar:1990qj,Brooks:2000nb},
providing important tests of the proton distribution amplitude.

\vspace{.5cm}

It is
particularly compelling to see a transition in angular dependence between
the low energy chiral and PQCD regimes.  The success of leading-twist
perturbative QCD scaling
for exclusive processes at presently experimentally accessible momentum
transfer can be understood if the effective coupling
$\alpha_V(Q^*)$ is approximately constant at the relatively
small scales $Q^*$ relevant to the hard scattering
amplitudes \cite{Brodsky:1998dh}.  The evolution of the quark distribution
amplitudes In the low-$Q^*$ domain at also needs to be minimal.  Sudakov
suppression of the endpoint contributions is also strengthened if the
coupling is frozen because of the exponentiation of a double logarithmic
series.

A debate has
continued \cite{Isgur:1989cy,Radyushkin:1998rt,Bolz:1996sw,Vogt:2000bz}
on whether processes such as the pion and
proton form factors and elastic Compton scattering $\gamma p \to \gamma
p$ might be dominated by higher-twist mechanisms until very large momentum
transfer.  If one
assumes that the light-front wavefunction of the pion has the form
$\psi_{\rm soft}(x,k_\perp) = A \exp (-b {k_\perp^2\over x(1-x)})$,
then the
Feynman endpoint contribution to the overlap integral at small $k_\perp$ and
$x \simeq 1$ will dominate the form factor compared to the hard-scattering
contribution until very large $Q^2$.  However, this ansatz for
$\psi_{\rm soft}(x,k_\perp)$ has no suppression at $k_\perp =0$
for any $x$; \ie, the
wavefunction in the hadron rest frame does not fall-off at all
for $k_\perp
= 0$ and $k_z \to - \infty$.  Thus such wavefunctions do not represent well
soft QCD contributions.  Endpoint
contributions are also suppressed by the QCD Sudakov form factor,
reflecting the fact that a near-on-shell quark must radiate if it absorbs
large momentum.
One can show \cite{Lepage:1980fj} that the leading
power dependence of the two-particle light-front Fock wavefunction in the
endpoint region is
$1-x$, giving a meson structure function which falls as $(1-x)^2$ and
thus by duality a non-leading contribution to the meson form factor
$F(Q^2)
\propto 1/Q^3$.  Thus the dominant contribution to meson form factors
comes from the hard-scattering regime.

Radyushkin \cite{Radyushkin:1998rt} has argued that the Compton amplitude is
dominated by soft
end-point contributions of the proton wavefunctions where the two photons
both interact on a quark line carrying nearly all of the
proton's momentum.  This description appears to agree with the Compton
data at least at forward angles where
$-t < 10$ GeV$^2$.  From this viewpoint, the dominance of the factorizable
PQCD leading twist contributions requires momentum transfers much
higher than those currently available.  However, the
endpoint model cannot explain the empirical success of the perturbative
QCD fixed $\theta_{c.m.}$ scaling $s^7 d\sigma/dt(\gamma p \to \pi^+ n)
\sim {\rm const} $
at relatively low momentum transfer in
pion photoproduction \cite{Anderson:1973cc}.

Clearly much more experimental input on hadron wavefunctions is needed,
particularly from measurements of two-photon exclusive reactions into
meson and baryon pairs at the high luminosity
$B$ factories.  For example, the ratio ${{d\sigma \over dt
}(\gamma
\gamma \to \pi^0
\pi^0)
/ {d\sigma \over dt}(\gamma \gamma \to \pi^+ \pi^-)}$
is particularly sensitive to the shape of pion distribution amplitude.
Baryon pair production in two-photon reactions at threshold may reveal
physics associated with the soliton structure of baryons in
QCD \cite{Sommermann:1992yh,Marek:2001}.  In addition, fixed target
experiments can provide much more information on fundamental QCD
processes such as deeply virtual Compton scattering and large angle
Compton scattering.

\section{Self-Resolved Diffractive Reactions and Light Cone Wavefunctions}

Diffractive multi-jet production in heavy
nuclei provides a novel way to measure the shape of the LC Fock
state wavefunctions and test color transparency.  For example, consider the
reaction \cite{Bertsch:1981py,Frankfurt:1993it,Frankfurt:2000tq}
$\pi A \rightarrow {\rm Jet}_1 + {\rm Jet}_2 + A^\prime$
at high energy where the nucleus $A^\prime$ is left intact in its ground
state.  The transverse momenta of the jets balance so that
$
\vec k_{\perp i} + \vec k_{\perp 2} = \vec q_\perp < {R^{-1}}_A \ .
$
The light-front longitudinal momentum fractions also need to add to
$x_1+x_2 \sim 1$ so that $\Delta p_L < R^{-1}_A$.  The process can
then occur coherently in the nucleus.

Because of color transparency, the
valence wavefunction of the pion with small impact separation, will
penetrate the nucleus with minimal interactions, diffracting into jet
pairs \cite{Bertsch:1981py}. The $x_1=x$, $x_2=1-x$ dependence of
the di-jet distributions will thus reflect the shape of the pion
valence light-front wavefunction in $x$; similarly, the
$\vec k_{\perp 1}- \vec k_{\perp 2}$ relative transverse momenta of the
jets gives key information on the derivative of the underlying shape
of the valence pion
wavefunction \cite{Frankfurt:1993it,Frankfurt:2000tq,BHDP}. The
diffractive nuclear amplitude extrapolated to
$t = 0$ should be linear in nuclear number $A$ if color transparency is
correct.  The integrated diffractive rate should then scale as $A^2/R^2_A
\sim A^{4/3}$.  Preliminary results on a diffractive dissociation
experiment of this type E791 at Fermilab using 500 GeV incident pions on
nuclear targets \cite{Aitala:2000hc} appear to be consistent with color
transparency.  The measured longitudinal momentum
distribution of the jets \cite{Aitala:2000hb} is consistent with a
pion light-cone wavefunction of the pion with the shape of the
asymptotic distribution amplitude,
$\phi^{\rm asympt}_\pi (x) =
\sqrt 3 f_\pi x(1-x)$.  Data from CLEO \cite{Gronberg:1998fj} for the
$\gamma \gamma^* \rightarrow \pi^0$ transition form factor also favor a form for
the pion distribution amplitude close to the asymptotic solution
to the perturbative QCD evolution
equation \cite{Lepage:1979zb,Lepage:1980fj}.

The diffractive dissociation of a hadron or nucleus can also occur via
the Coulomb dissociation of a beam particle on an electron beam (\eg\ at
HERA or eRHIC) or on the strong Coulomb field of a heavy nucleus (\eg\
at RHIC or nuclear collisions at the LHC) \cite{BHDP}. The amplitude for
Coulomb exchange at small momentum transfer is proportional to the first
derivative $\sum_i e_i {\partial \over \vec k_{T i}} \psi$ of the
light-front wavefunction, summed over the charged constituents.  The Coulomb
exchange reactions fall off less fast at high transverse momentum compared
to pomeron exchange reactions since the light-front wavefunction is
effective differentiated twice in two-gluon exchange reactions.

It will also be interesting to study diffractive tri-jet production
using proton beams
$ p A \rightarrow {\rm Jet}_1 + {\rm Jet}_2 + {\rm Jet}_3 + A^\prime $ to
determine the fundamental shape of the 3-quark structure of the valence
light-front wavefunction of the nucleon at small transverse
separation \cite{Frankfurt:1993it}.
For example, consider the Coulomb dissociation of a high energy proton at
HERA.  The proton can dissociate into three jets corresponding to the
three-quark structure of the valence light-front wavefunction.  We can
demand that the produced hadrons all fall outside an opening angle $\theta$
in the proton's fragmentation region.
Effectively all of the light-front momentum
$\sum_j x_j \simeq 1$ of the proton's fragments will thus be
produced outside an ``exclusion cone".  This
then limits the invariant mass of the contributing Fock state ${\cal
M}^2_n >
\Lambda^2 = P^{+2} \sin^2\theta/4$ from below, so that perturbative QCD
counting rules can predict the fall-off in the jet system invariant mass
$\cal M$.  At large invariant mass one expects the three-quark valence
Fock state of the proton to dominate.  The segmentation of the forward
detector in azimuthal angle $\phi$ can be used to identify structure and
correlations associated with the three-quark light-front
wavefunction \cite{BHDP}.
An interesting possibility is
that the distribution amplitude of the
$\Delta(1232)$ for $J_z = 1/2, 3/2$ is close to the asymptotic form $x_1
x_2 x_3$,  but that the proton distribution amplitude is more complex.
This ansatz can also be motivated by assuming a quark-diquark structure
of the baryon wavefunctions.  The differences in shapes of the
distribution amplitudes could explain why the $p
\to\Delta$ transition form factor appears to fall faster at large $Q^2$
than the elastic $p \to p$ and the other $p \to N^*$ transition form
factors \cite{Stoler:1999nj}. One can use also measure the dijet
structure of real and virtual photons beams
$ \gamma^* A \rightarrow {\rm Jet}_1 + {\rm Jet}_2 + A^\prime $ to
measure the shape of the light-front wavefunction for
transversely-polarized and longitudinally-polarized virtual photons.  Such
experiments will open up a direct window on the amplitude
structure of hadrons at short distances.
The light-front formalism is also applicable to the
description of nuclei in terms of their nucleonic and mesonic
degrees of freedom \cite{Miller:1999mi,Miller:2000ta}.
Self-resolving diffractive jet reactions
in high energy electron-nucleus collisions and hadron-nucleus collisions
at moderate momentum transfers can thus be used to resolve the light-front
wavefunctions of nuclei.

\section{Higher Fock States and the Intrinsic Sea}

One can identify two contributions to the
heavy quark sea, the ``extrinsic'' contributions which correspond to
ordinary gluon splitting, and the ``intrinsic" sea which is
multi-connected via gluons to the valence quarks.  The leading $1/m_Q^2$
contributions to the intrinsic sea of the proton in the heavy quark
expansion are proton matrix elements of the operator~ \cite{Franz:2000ee}
$\eta^\mu \eta^\nu G_{\alpha \mu} G_{\beta \nu} G^{\alpha \beta}$ which
in light-cone gauge $\eta^\mu A_\mu= A^+= 0$ corresponds to three or four
gluon exchange between the heavy-quark loop and the proton constituents
in the forward virtual Compton amplitude.  The intrinsic sea is thus
sensitive to the hadronic bound-state
structure \cite{Brodsky:1981se,Brodsky:1980pb}. The maximal contribution
of the intrinsic heavy quark occurs at $x_Q \simeq {m_{\perp Q}/ \sum_i
m_\perp}$ where
$m_\perp = \sqrt{m^2+k^2_\perp}$;
\ie\ at large $x_Q$, since this minimizes the invariant mass $\M^2_n$.
The
measurements of the charm structure function by the EMC experiment are
consistent with intrinsic charm at large $x$ in the nucleon with a
probability of order $0.6 \pm 0.3 \% $ \cite{Harris:1996jx} which is
consistent with the recent estimates based on instanton
fluctuations \cite{Franz:2000ee}.

Chang and Hou \cite{Chang:2001iy} have recently discussed the consequences of
intrinsic charm in heavy quark states such as the $B$, $\Lambda_B$, and
$\Upsilon$, such as an anomalous momentum distribution for
$B \to J/\psi X$.   The characteristic momenta characterizing the $B$ meson is
most likely higher by a factor of 2 compared to the momentum scale of light
mesons, This effect is analogous to the higher momentum scale of muonium
$\mu^+ e^-$ versus
that of positronium $e^+ e^-$ in atomic physics because of the larger reduced
mass.  Thus one can expect a higher probability for intrinsic charm in
heavy hadrons compared to light hadrons.

One can also distinguish
``intrinsic gluons" \cite{Brodsky:1990db} which are associated with
multi-quark interactions and extrinsic gluon contributions associated
with quark substructure.  One can also use this framework to
isolate the physics of the anomaly contribution to the Ellis-Jaffe sum
rule \cite{Bass:1999rn}. Thus neither gluons nor sea quarks are solely
generated by DGLAP evolution, and one cannot define a resolution scale
$Q_0$ where the sea or gluon degrees of freedom can be neglected.

It is usually assumed that a heavy quarkonium state such as the
$J/\psi$ always decays to light hadrons via the annihilation of its heavy quark
constituents to gluons.  However, as Karliner and I \cite{Brodsky:1997fj}
have shown, the transition $J/\psi \to \rho
\pi$ can also occur by the rearrangement of the $c \bar c$ from the $J/\psi$
into the $\ket{ q \bar q c \bar c}$ intrinsic charm Fock state of the $\rho$ or
$\pi$.  On the other hand, the overlap rearrangement integral in the
decay $\psi^\prime \to \rho \pi$ will be suppressed since the intrinsic
charm Fock state radial wavefunction of the light hadrons will evidently
not have nodes in its radial wavefunction.  This observation provides
a natural explanation of the long-standing puzzle~ \cite{Brodsky:1987bb}
why the $J/\psi$ decays prominently to two-body pseudoscalar-vector final
states, breaking hadron helicity
conservation~ \cite{Brodsky:1981kj}, whereas the
$\psi^\prime$ does not.

The higher Fock state of the proton $\ket{u u d s \bar s}$ should
resemble a $\ket{ K \Lambda}$ intermediate state, since this minimizes its
invariant mass $\M$.  In such a state, the
strange quark has a higher mean momentum fraction $x$ than the $\bar
s$ \cite{Burkardt:1992di,Signal:1987gz,Brodsky:1996hc}.  Similarly, the
helicity of the intrinsic strange quark in this configuration will be
anti-aligned with the helicity of the
nucleon \cite{Burkardt:1992di,Brodsky:1996hc}.  This $Q
\leftrightarrow
\bar Q$ asymmetry is a striking feature of the intrinsic heavy-quark sea.

\section{Non-Perturbative Solutions of Light-Front Quantized QCD}

Is there any hope of computing light-front wavefunctions from
first principles?  The solution of the light-front Hamiltonian equation
$ H^{QCD}_{LC}
\ket{\Psi} = M^2 \ket{\Psi}$ is an eigenvalue problem which in principle
determines the masses squared of the entire bound and continuum spectrum
of QCD.  If one introduces periodic or anti-periodic boundary conditions,
the eigenvalue problem is reduced to the diagonalization of a
discrete Hermitian matrix representation of $H^{QCD}_{LC}.$ The
light-front momenta satisfy
$x^+ = {2
\pi
\over L} n_i$ and
$P^+ = {2\pi \over L} K$, where $\sum_i n_i = K.$ The number of quanta in
the contributing Fock states is restricted by the choice of harmonic
resolution.  A cutoff on the invariant mass of the Fock states truncates
the size of the matrix representation in the transverse momenta.
This is
the essence of the DLCQ method \cite{Pauli:1985ps}, which has now become
a standard tool for solving both the spectrum and light-front
wavefunctions
of one-space one-time theories -- virtually any
$1+1$ quantum field theory, including ``reduced QCD"
(which has both quark and
gluonic degrees of freedom) can be completely solved
using
DLCQ \cite{Dalley:1993yy,Antonuccio:1995fs}.
The method yields not only the bound-state and continuum
spectrum, but also the light-front wavefunction
for each eigensolution \cite{Antonuccio:1996hv,Antonuccio:1996rb}.

In the case of theories in 3+1 dimensions, Hiller, McCartor, and I\
 \cite{Brodsky:1998hs,Brodsky:1999xj} have recently shown that the use of
covariant Pauli-Villars regularization with DLCQ allows one to obtain the
spectrum and light-front wavefunctions of simplified theories, such as
(3+1) Yukawa theory.  Dalley \etal\ have shown how one can use DLCQ
in one space-one time, with a transverse lattice to solve mesonic
and gluonic states in $ 3+1$ QCD \cite{Dalley:2000ii}. The spectrum
obtained for gluonium states
is in remarkable agreement with lattice gauge theory results, but with a
huge reduction of numerical effort.  Hiller and I \cite{Hiller:1999cv}
have shown how one can use DLCQ to compute the electron magnetic moment
in QED without resort to perturbation theory.

One can also formulate DLCQ so
that supersymmetry is exactly preserved in the discrete approximation,
thus combining the power of DLCQ with the beauty of
supersymmetry \cite{Antonuccio:1999ia,Lunin:1999ib,Haney:2000tk}. The
``SDLCQ" method has been applied to several interesting supersymmetric
theories, to the analysis of zero modes, vacuum degeneracy, massless
states, mass gaps, and theories in higher dimensions, and even tests of
the Maldacena conjecture \cite{Antonuccio:1999ia}.
Broken supersymmetry is
interesting in DLCQ, since it may serve as a method for regulating
non-Abelian theories \cite{Brodsky:1999xj}.

There are also many possibilities for obtaining approximate solutions of
light-front wavefunctions in QCD.  QCD sum rules, lattice gauge theory
moments, and QCD inspired models
such as the bag model, chiral theories, provide important constraints.
Guides to the exact behavior of LC wavefunctions in
QCD can also be obtained from analytic or DLCQ solutions to toy models
such as ``reduced"
$QCD(1+1).$ The light-front and many-body
Schr\"odinger theory formalisms must match In the nonrelativistic limit.

It would be interesting to see if light-front wavefunctions can
incorporate chiral constraints such as soliton (Skyrmion) behavior for
baryons and other consequences of the chiral limit in the soft momentum
regime.  Solvable theories such as $QCD(1+1)$ are also useful for
understanding such phenomena.  It has been shown that the anomaly
contribution for the $\pi^0\to \gamma \gamma$ decay amplitude is
satisfied by the light-front Fock formalism in the limit where the mass
of the pion is light compared to its size \cite{Lepage:1982gd}.

\section{Non-Perturbative Calculations of the Pion Distribution Amplitude}

The distribution amplitude $\phi(x,\widetilde Q)$ can be computed from
the integral over transverse momenta of the renormalized hadron valence
wavefunction in the light-cone gauge at fixed light-cone
time \cite{Brodsky:1989pv}:
\begin{equation}
\phi(x,\widetilde Q) = \int d^2\vec{k_\perp}\thinspace
\theta \left({\widetilde Q}^2 - {\vec{k_\perp}^2\over x(1-x)}\right)
\psi^{(\widetilde Q)}(x,\vec{k_\perp}),
\label{quarkdistamp}
\end{equation}
where a global cutoff in invariant mass is identified with the resolution
$\tilde Q$.  The distribution amplitude $\phi(x, \tilde Q)$ is boost and
gauge invariant and evolves in $\ln \tilde Q$ through an evolution
equation \cite{Lepage:1979za,Lepage:1979zb,Lepage:1980fj}.
Since it is formed
from the same product of operators as the non-singlet structure function,
the anomalous dimensions controlling $\phi(x,Q)$ dependence in the
ultraviolet
$\log Q$ scale are the same as those which appear in the DGLAP
evolution of structure functions \cite{Brodsky:1980ny}.
The
decay $\pi \to \mu \nu$ normalizes the wave function at the origin:
${a_0/ 6} = \int^1_0 dx \phi(x,Q) =
{f_\pi/ (2 \sqrt 3)}.$
One can also compute the distribution amplitude from the gauge invariant
Bethe-Salpeter wavefunction at equal light-cone time.  This also allows
contact with both QCD sum rules
and lattice
gauge theory; for example, moments of the pion distribution amplitudes
have been computed in lattice
gauge theory \cite{Martinelli:1987si,Daniel:1991ah,DelDebbio:2000mq}.

\begin{figure}[htb]
\begin{center}
\leavevmode
{\epsfxsize=5.5in\epsfbox{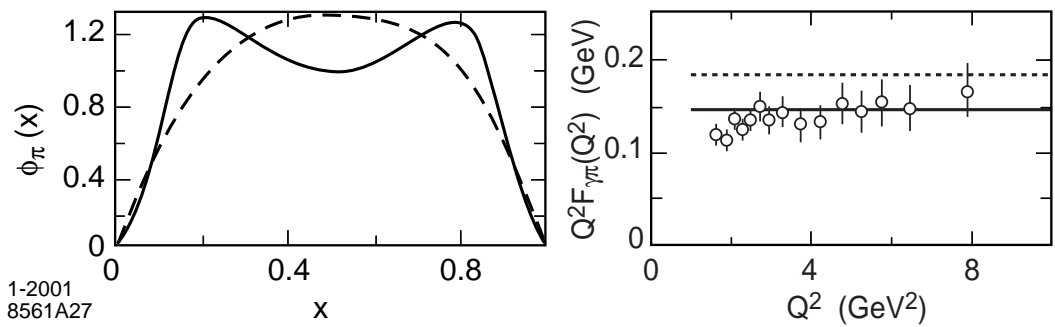}}
\end{center}
\caption[*]{
(a) Preliminary transverse lattice results for the pion distribution
amplitude at $Q^2 \sim 10 ~{\rm GeV}^2$.  The solid curve is the theoretical
prediction from the combined DLCQ/transverse lattice
method \cite{Dalley:2000dh}; the chain line is the experimental result
obtained from jet diffractive dissociation \cite{Ashery:1999nq}.  Both are
normalized to the same area for comparison.
(b) Scaling of the transition photon to pion transition form
factor $Q^2F_{\gamma \pi^0}(Q^2)$.  The dotted and
solid theoretical curves are the perturbative QCD prediction at leading
and next-to-leading order, respectively, assuming the asymptotic pion
distribution The data are from the CLEO
collaboration \cite{Gronberg:1998fj}.}
\label{Fig:DalleyCleo}
\end{figure}

Dalley \cite{Dalley:2000dh} has recently calculated the pion
distribution amplitude from QCD using a combination of the discretized
DLCQ method for the $x^-$
and
$x^+$ light-cone coordinates with the transverse
lattice method \cite{Bardeen:1976tm,Burkardt:1996gp} in the transverse
directions,  A finite lattice spacing $a$ can be used by choosing
the parameters of the effective theory in a region of
renormalization group stability to respect the required gauge,
Poincar\'e, chiral, and continuum symmetries.
The overall normalization gives $f_{\pi} = 101$ MeV
compared with the experimental value of $93$ MeV.
Figure \ref{Fig:DalleyCleo} (a)
compares the resulting DLCQ/transverse lattice pion wavefunction with
the best fit to the diffractive di-jet data (see the next section) after
corrections for hadronization and experimental acceptance
\cite{Ashery:1999nq}.
The theoretical curve
is somewhat broader than the experimental result.  However, there
are experimental uncertainties from hadronization and
theoretical errors introduced from finite DLCQ resolution,
using a nearly massless pion, ambiguities in setting the factorization
scale
$Q^2$, as well as errors in the evolution of the distribution amplitude
from 1 to $10~{\rm GeV}^2$.  Instanton models also predict a pion
distribution
amplitude close to the asymptotic form \cite{Petrov:1999kg}.
In contrast,  recent lattice results from Del Debbio
{\em et al.} \cite{DelDebbio:2000mq} predict a much narrower
shape for the pion distribution amplitude than the distribution predicted
by the transverse lattice.
A new result for the proton distribution amplitude treating nucleons as
chiral solitons
has recently been derived by Diakonov and Petrov \cite{Diakonov:2000pa}.
Dyson-Schwinger models \cite{Hecht:2000xa} of
hadronic Bethe-Salpeter wavefunctions can also be used to
predict light-cone wavefunctions and hadron distribution amplitudes by
integrating over the relative $k^-$ momentum.  There is also the
possibility of deriving Bethe-Salpeter wavefunctions within light-cone
gauge quantized QCD \cite{Srivastava:2000gi} in order to properly match to
the light-cone gauge Fock state decomposition.

\vspace{.5cm}

\section{Conclusions}

In these lectures I have
shown how the universal, process-independent and
frame-independent light-front Fock-state wavefunctions
encode the properties of a hadron in terms of its fundamental quark and
gluon degrees of freedom. Knowledge of such wavefunctions  will be
critical for progress in understanding exclusive $B$ decays.

I have shown how, given the proton's light-front wavefunctions, one
can compute not only the moments of the quark and gluon distributions
measured in deep inelastic lepton-proton scattering, but also the
multi-parton correlations which control the distribution of particles in
the proton fragmentation region and dynamical higher twist effects.
Light-front wavefunctions also provide a systematic framework for
evaluating exclusive hadronic matrix elements, including time-like heavy
hadron decay amplitudes,  form factors, and the generalized form factors
that appear in deeply virtual Compton scattering. The light-front
Hamiltonian  formalism also provides a physical factorization scheme for
separating hard and soft contributions
in both exclusive and inclusive hard processes.

The leading-twist QCD predictions for exclusive two-photon processes such
as the photon-to-pion transition form factor and $\gamma \gamma \to $
hadron pairs are based on rigorous factorization theorems.  The recent
data from the CLEO collaboration on $F_{\gamma \pi}(Q^2)$ and the sum of
$\gamma \gamma \to \pi^+ \pi^-$ and $\gamma \gamma \to K^+ K^-$
channels are in excellent agreement with the QCD predictions.  It is
particularly compelling to see a transition in angular dependence between
the low energy chiral and PQCD regimes.  The success of leading-twist
perturbative QCD scaling
for exclusive processes at presently experimentally accessible momentum
transfer can be understood if the effective coupling
$\alpha_V(Q^*)$ is approximately constant at the relatively
small scales $Q^*$ relevant to the hard scattering
amplitudes \cite{Brodsky:1998dh}. The evolution of the quark distribution
amplitudes In the low-$Q^*$ domain at also needs to be minimal.  Sudakov
suppression of the endpoint contributions is also strengthened if the
coupling is frozen because of the exponentiation of a double logarithmic
series.

One of the formidable challenges in QCD is the calculation of
non-perturbative wavefunctions of hadrons from first principles.  The
recent calculation of the pion distribution amplitude by
Dalley \cite{Dalley:2000dh} using light-cone and transverse lattice
methods is particularly encouraging.  The predicted form of
$\phi_\pi(x,Q)$ is somewhat broader than but not inconsistent with the
asymptotic form favored by the measured normalization of $Q^2 F_{\gamma
\pi^0}(Q^2)$ and the pion wavefunction inferred from diffractive di-jet
production.

Clearly much more experimental input on hadron wavefunctions is needed,
particularly from measurements of two-photon exclusive reactions into
meson and baryon pairs at the high luminosity
$B$ factories.  For example, the ratio ${{d\sigma \over dt
}(\gamma
\gamma \to \pi^0
\pi^0)
/ {d\sigma \over dt}(\gamma \gamma \to \pi^+ \pi^-)}$
is particularly sensitive to the shape of pion distribution amplitude.
Baryon pair production in two-photon reactions at threshold may reveal
physics associated with the soliton structure of baryons in
QCD \cite{Sommermann:1992yh}. In addition, fixed target experiments can
provide much more information on fundamental QCD processes such as deeply
virtual Compton scattering and large angle Compton scattering.

A remarkable new type of diffractive jet
production reaction, ``self-resolving diffractive interactions" can
provide direct empirical information on the light-front wavefunctions of
hadrons.
The recent
E791 experiment at Fermilab has not only determined the main features of
the pion wavefunction, but has also
confirmed color transparency, a fundamental test of the gauge properties of
QCD.
Analogous reaction involving nuclear projectiles can resolve the light-front
wavefunctions of
nuclei in terms of their nucleon and mesonic degrees of freedom.
It is also possible to measure the light-front wavefunctions of atoms through
high energy Coulomb dissociation.

There has been notable progress in computing
light-front wavefunctions directly from the QCD light-front Hamiltonian,
using DLCQ and transverse lattice methods. Even without full
non-perturbative solutions of QCD, one can envision a
program to construct the light-front wavefunctions using measured moments
constraints from QCD sum rules, lattice gauge theory, and data from
hard exclusive and
inclusive processes.  One can also be guided by theoretical constraints from
perturbation theory which dictate the asymptotic form of the
wavefunctions at large invariant mass,
$x \to 1$, and high
$k_\perp$.  One can also use ladder relations which connect Fock states of
different particle number; perturbatively-motivated numerator spin
structures; conformal symmetry, guidance from toy models
such as ``reduced"
$QCD(1+1)$; and the correspondence to Abelian theory
for
$N_C\to 0$, as well as many-body
Schr\"odinger theory in the nonrelativistic domain.

\section*{Acknowledgments}
Work supported by the Department of Energy
under contract number DE-AC03-76SF00515.
I wish to thank the organizers of this meeting, Hsiang-nan Li and
Wei-Min Zhang,  for their outstanding hospitality in Taiwan.  Much of this
work is based on collaborations, particularly with Markus Diehl, Paul
Hoyer, Dae Sung Hwang,  Peter Lepage, Bo-Qiang Ma,  Hans Christian Pauli,
Johan Rathsman,  Ivan Schmidt, and Prem Srivastava.


\begin{thebibliography}{99}

\bibitem{Ashery:1999nq}
D.~Ashery [E791 Collaboration],
hep-ex/9910024.


\bibitem{Bertsch:1981py}
G.~Bertsch, S.~J.~Brodsky, A.~S.~Goldhaber and J.~F.~Gunion,
Phys.\ Rev.\ Lett.\ {\bf 47}, 297 (1981).


\bibitem{Frankfurt:1993it}
L.~Frankfurt, G.~A.~Miller and M.~Strikman,
Phys.\ Lett.\ {\bf B304}, 1 (1993)
[hep-ph/9305228].

\bibitem{Frankfurt:2000tq}
L.~Frankfurt, G.~A.~Miller and M.~Strikman,
Found.\ Phys.\ {\bf 30}, 533 (2000)
[hep-ph/9907214].

\bibitem{Lepage:1980fj}
G.~P.~Lepage and S.~J.~Brodsky,
Phys.\ Rev.\ D {\bf 22}, 2157 (1980).

\bibitem{Beneke:1999br}
M.~Beneke, G.~Buchalla, M.~Neubert and C.~T.~Sachrajda,
Phys.\ Rev.\ Lett.\ {\bf 83}, 1914 (1999)
[hep-ph/9905312].

\bibitem{Keum:2000ph}
Y.~Keum, H.~Li and A.~I.~Sanda,
hep-ph/0004004.

\bibitem{Signal:1987gz}
A.~I.~Signal and A.~W.~Thomas,
Phys.\ Lett.\ {\bf B191}, 205 (1987).

\bibitem{Martinelli:1987si}
G.~Martinelli and C.~T.~Sachrajda,
Amplitude,''
Phys.\ Lett.\ {\bf B190}, 151 (1987).

\bibitem{Hecht:2000xa}
M.~B.~Hecht, C.~D.~Roberts and S.~M.~Schmidt,
nucl-th/0008049.


\bibitem{Robertson:1999va}
D.~G.~Robertson, E.~S.~Swanson, A.~P.~Szczepaniak, C.~R.~Ji and S.~R.~Cotanch,
Phys.\ Rev.\ D {\bf 59}, 074019 (1999)
[hep-ph/9811224].

\bibitem{Brodsky:1998de}
S.~J.~Brodsky, H.~Pauli and S.~S.~Pinsky,
Phys.\ Rept.\ {\bf 301}, 299 (1998)
[hep-ph/9705477].


\bibitem{Dirac:1949cp}
P.~A.~Dirac,
Rev.\ Mod.\ Phys.\  {\bf 21}, 392 (1949).


\bibitem{Weinberg:1966jm}
S.~Weinberg,
Phys.\ Rev.\ {\bf 150}, 1313 (1966).


\bibitem{Brodsky:1973kb}
S.~J.~Brodsky, R.~Roskies and R.~Suaya,
Momentum Frame,''
Phys.\ Rev.\ D {\bf 8}, 4574 (1973).

\bibitem{Brodsky:1989pv}
S. J. Brodsky and G. P. Lepage, in {\em Perturbative Quantum
Chromodynamics}, A. H. Mueller, Ed.  (World Scientific, 1989).

\bibitem{Brodsky:2001ii}
S.~J.~Brodsky, D.~S.~Hwang, B.~Ma and I.~Schmidt,
Nucl.\ Phys.\ {\bf B593}, 311 (2001)
[hep-th/0003082].

\bibitem{Srivastava:2000cf}
P.~P.~Srivastava and S.~J.~Brodsky,
hep-ph/0011372.

\bibitem{Cornwall:1989gv}
J.~M.~Cornwall and J.~Papavassiliou,
Phys.\ Rev.\ D {\bf 40}, 3474 (1989).


\bibitem{Brodsky:2000cr}
S.~J.~Brodsky, E.~Gardi, G.~Grunberg and J.~Rathsman,
hep-ph/0002065.

\bibitem{Srivastava:2000gi}
P.~P.~Srivastava and S.~J.~Brodsky,
Phys.\ Rev.\  {\bf D61}, 025013 (2000), \hfill\break
hep-ph/9906423, and SLAC-PUB 8543, in preparation.


\bibitem{Bassetto:1999tm}
A.~Bassetto, L.~Griguolo and F.~Vian,
hep-th/9911036.


\bibitem{Yamawaki:1998cy}
K.~Yamawaki,
hep-th/9802037.


\bibitem{McCartor:2000yy}
G.~McCartor,
hep-th/0004139.


\bibitem{Srivastava:1999et}
P.~P.~Srivastava,
Phys.\ Lett.\ {\bf B448}, 68 (1999)
[hep-th/9811225].


\bibitem{Pinsky:1994yi}
S.~S.~Pinsky and B.~van de Sande,
Phys.\ Rev.\  {\bf D49}, 2001 (1994),
hep-ph/9310330.

\bibitem{Drell:1970km}
S.~D.~Drell and T.~Yan,
Phys.\ Rev.\ Lett.\ {\bf 24}, 181 (1970).

\bibitem{West:1970av}
G.~B.~West,
Phys.\ Rev.\ Lett.\ {\bf 24}, 1206 (1970).

\bibitem{Brodsky:1980zm}
S.~J.~Brodsky and S.~D.~Drell,
Phys.\ Rev.\ D {\bf 22}, 2236 (1980).

\bibitem{Brodsky:1999hn}
S.~J.~Brodsky and D.~S.~Hwang,
Nucl.\ Phys.\ {\bf B543}, 239 (1999)
[hep-ph/9806358].


\bibitem{Brodsky:2000xy}
S.~J.~Brodsky, M.~Diehl and D.~S.~Hwang,
hep-ph/0009254.

\bibitem{Diehl:2000xz}
M.~Diehl, T.~Feldmann, R.~Jakob and P.~Kroll,
hep-ph/0009255.


\bibitem{Brodsky:1995kg}
S.~J.~Brodsky, M.~Burkardt and I.~Schmidt,
distributions,''
Nucl.\ Phys.\ {\bf B441}, 197 (1995)
[hep-ph/9401328].



\bibitem{Brodsky:2000zu}
S.~J.~Brodsky,
hep-ph/0006310.

\bibitem{Brodsky:1980pb}
S.~J.~Brodsky, P.~Hoyer, C.~Peterson and N.~Sakai,
Phys.\ Lett.\ {\bf B93}, 451 (1980).

\bibitem{Harris:1996jx}
B.~W.~Harris, J.~Smith and R.~Vogt,
Nucl.\ Phys.\ {\bf B461}, 181 (1996)
[hep-ph/9508403].

\bibitem{Antonuccio:1997tw}
F.~Antonuccio, S.~J.~Brodsky and S.~Dalley,
Phys.\ Lett.\ {\bf B412}, 104 (1997)
[hep-ph/9705413].

\bibitem{Brodsky:1988xz}
S.~J.~Brodsky and A.~H.~Mueller,
Phys.\ Lett.\ {\bf B206}, 685 (1988).



\bibitem{Frankfurt:1988nt}
L.~L.~Frankfurt and M.~I.~Strikman,
Phys.\ Rept.\ {\bf 160}, 235 (1988).



\bibitem{Brodsky:1976rz}
S.~J.~Brodsky and B.~T.~Chertok,
Phys.\ Rev.\ D {\bf 14}, 3003 (1976).

\bibitem{Brodsky:1983vf}
S.~J.~Brodsky, C.~Ji and G.~P.~Lepage,
Phys.\ Rev.\ Lett.\  {\bf 51}, 83 (1983).


\bibitem{Farrar:1991qi}
G. R.~Farrar, K.~Huleihel and H.~Zhang,
{Phys. Rev. Lett.} {\bf 74}, 650 (1995).


\bibitem{Brodsky:2000zc}
S.~J.~Brodsky, E.~Chudakov, P.~Hoyer and J.~M.~Laget,
hep-ph/0010343.

\bibitem{Brodsky:1999gm}
S. J.~Brodsky and J.~Rathsman, hep-ph/9906339.


\bibitem{Brodsky:1980ny}
S. J.~Brodsky, Y.~Frishman, G. P.~Lepage and C.~Sachrajda,
{\em Phys. Lett.} {\bf 91B}, 239 (1980).

\bibitem{Muller:1994hg}
D.~M\"uller, {\em Phys. Rev.} {\bf D49}, 2525 (1994).

\bibitem{Braun:1999te}
V.~M.~Braun, S.~E.~Derkachov, G.~P.~Korchemsky and A.~N.~Manashov,
Nucl.\ Phys.\  {\bf B553}, 355 (1999),
hep-ph/9902375.


\bibitem{Brodsky:1995eh}
S. J.~Brodsky and H. J.~Lu,
{\em Phys. Rev.} {\bf D51}, 3652 (1995),
hep-ph/9405218.

\bibitem{Brodsky:1998ua}
S. J.~Brodsky, J. R.~Pelaez and N.~Toumbas,
Phys.\ Rev.\ {\bf D60}, 037501 (1999),
hep-ph/9810424.


\bibitem{Brodsky:1996tb}
S. J.~Brodsky, G. T.~Gabadadze, A. L.~Kataev and H. J.~Lu,
{\em Phys. Lett.} {\bf B372}, 133 (1996) hep-ph/9512367.


\bibitem{Brodsky:1998mf}
S.~J.~Brodsky, M.~S.~Gill, M.~Melles and J.~Rathsman,
Phys.\ Rev.\ D {\bf 58}, 116006 (1998)
[hep-ph/9801330].

\bibitem{Brodsky:1997jk}
S. J.~Brodsky and P.~Huet, Phys. Lett. {\bf B417}, 145 (1998),
hep-ph/9707543.


\bibitem{Okun}
L.~Okun and  I.~Yu.~Kobzarev,
ZhETF, {\bf 43} 1904 (1962)  ( English
translation : JETP {\bf 16}  1343 (1963));
L.~Okun, in proceedings of the
International Conference on Elementary Particles, 4th, Heidelberg,
Germany (1967). Edited by H.~Filthuth. North-Holland, (1968).

\bibitem{Ji:1996kb}
X.~Ji,
hep-ph/9610369.

\bibitem{Ji:1997ek}
X.~Ji,
Phys.\ Rev.\ Lett.\  {\bf 78}, 610 (1997), hep-ph/9603249.



\bibitem{Ji:1997nm}
X.~Ji,
Phys.\ Rev.\  {\bf D55}, 7114 (1997), hep-ph/9609381.

\bibitem{Teryaev:1999su}
O.~V.~Teryaev,
hep-ph/9904376.

\bibitem{Harindranath:1999ve}
A.~Harindranath and R.~Kundu,
Phys.\ Rev.\  {\bf D59}, 116013 (1999),\hfill \break
hep-ph/9802406.

\bibitem{Chang:1973xt}
S.~Chang, R.~G.~Root and T.~Yan,
Phys.\ Rev.\ D {\bf 7}, 1133 (1973).

\bibitem{Burkardt:1989wy}
M.~Burkardt,
Nucl.\ Phys.\ {\bf A504}, 762 (1989).

\bibitem{Choi:1998nf}
H.-M.~Choi and C.-R.~Ji,
Phys. Rev. {\bf D 58}, 071901 (1998). 

\bibitem{Hornbostel:1990fb}
K.~Hornbostel, S.~J.~Brodsky and H.~C.~Pauli,
Phys.\ Rev.\ D {\bf 41}, 3814 (1990).

\bibitem{Antonuccio:1995fs}
F.~Antonuccio and S.~Dalley,
Phys.\ Lett.\ {\bf B348}, 55 (1995)
[hep-th/9411204].

\bibitem{Brodsky:1972zh}
S.~J.~Brodsky, F.~E.~Close and J.~F.~Gunion,
Phys.\ Rev.\ D {\bf 5}, 1384 (1972).

\bibitem{Brodsky:1972vv}
S.~J.~Brodsky, F.~E.~Close and J.~F.~Gunion,
Phys.\ Rev.\ D {\bf 6}, 177 (1972).

\bibitem{Brodsky:1973hm}
S.~J.~Brodsky, F.~E.~Close and J.~F.~Gunion,
Phys.\ Rev.\ D {\bf 8}, 3678 (1973).

\bibitem{Diehl:2000uv}
M.~Diehl, T.~Gousset and B.~Pire,
Phys.\ Rev.\ D {\bf 62}, 073014 (2000)
[hep-ph/0003233].


\bibitem{Radyushkin:1996nd}
A.~V.~Radyushkin,
Phys.\ Lett.\ {\bf B380}, 417 (1996)
[hep-ph/9604317].





\bibitem{Ji:1998xh}
X.~Ji and J.~Osborne,
Phys.\ Rev.\ D {\bf 58}, 094018 (1998)
[hep-ph/9801260].

\bibitem{Guichon:1998xv}
P.~A.~Guichon and M.~Vanderhaeghen,
Prog.\ Part.\ Nucl.\ Phys.\ {\bf 41}, 125 (1998)
[hep-ph/9806305].

\bibitem{Vanderhaeghen:1998uc}
M.~Vanderhaeghen, P.~A.~Guichon and M.~Guidal,
Phys.\ Rev.\ Lett.\ {\bf 80}, 5064 (1998).

\bibitem{Radyushkin:1999es}
A.~V.~Radyushkin,
Phys.\ Rev.\ D {\bf 59}, 014030 (1999)
[hep-ph/9805342].

\bibitem{Collins:1999be}
J.~C.~Collins and A.~Freund,
Phys.\ Rev.\ D {\bf 59}, 074009 (1999)
[hep-ph/9801262].

\bibitem{Diehl:1999tr}
M.~Diehl, T.~Feldmann, R.~Jakob and P.~Kroll,
Phys.\ Lett.\ {\bf B460}, 204 (1999)
[hep-ph/9903268].

\bibitem{Diehl:1999kh}
M.~Diehl, T.~Feldmann, R.~Jakob and P.~Kroll,
Eur.\ Phys.\ J.\ {\bf C8}, 409 (1999)
[hep-ph/9811253].

\bibitem{Blumlein:2000cx}
J.~Blumlein and D.~Robaschik,
Nucl.\ Phys.\ {\bf B581}, 449 (2000)
[hep-ph/0002071].

\bibitem{Penttinen:2000dg}
M.~Penttinen, M.~V.~Polyakov, A.~G.~Shuvaev and M.~Strikman,
Phys.\ Lett.\ {\bf B491}, 96 (2000)
[hep-ph/0006321].



\bibitem{Muller:1994fv}
D.~Muller, D.~Robaschik, B.~Geyer, F.~M.~Dittes and J.~Horejsi,
Fortsch.\ Phys.\  {\bf 42}, 101 (1994)
[hep-ph/9812448].

\bibitem{Brodsky:2000dr}
S.~J.~Brodsky,
SLAC-PUB-8649.


\bibitem{Lepage:1979zb}
G. P.~Lepage and S. J.~Brodsky,
Phys.\ Lett.\  {\bf B 87}, 359 (1979).


\bibitem{Bass:1999rn}
S.~D.~Bass, S.~J.~Brodsky and I.~Schmidt,
Phys.\ Rev.\ D {\bf 60}, 034010 (1999)
[hep-ph/9901244].

\bibitem{Brodsky:1975vy}
S.~J.~Brodsky and G.~R.~Farrar,
Phys.\ Rev.\ D {\bf 11}, 1309 (1975).


\bibitem{Lepage:1979za}
G.~P.~Lepage and S.~J.~Brodsky,
Phys.\ Rev.\ Lett.\  {\bf 43}, 545 (1979).

\bibitem{Szczepaniak:1990dt}
A.~Szczepaniak, E.~M.~Henley and S.~J.~Brodsky,
Phys.\ Lett.\ {\bf B243}, 287 (1990).



\bibitem{Szczepaniak:1996xg}
A.~Szczepaniak,
Phys.\ Rev.\ D {\bf 54}, 1167 (1996).

\bibitem{Keum:2000wi}
Y.~Y.~Keum, H.~Li and A.~I.~Sanda,
hep-ph/0004173.

\bibitem{Li:2000hh}
H.~Li,
hep-ph/0012140.

\bibitem{Brodsky:1981kj}
S.~J.~Brodsky and G.~P.~Lepage,
Phys.\ Rev.\ D {\bf 24}, 2848 (1981).


\bibitem{Brodsky:1998dh}
S.~J.~Brodsky, C.~Ji, A.~Pang and D.~G.~Robertson,
Phys.\ Rev.\ D {\bf 57}, 245 (1998)
[hep-ph/9705221].

\bibitem{Brodsky:1981rp}
S.~J.~Brodsky and G.~P.~Lepage,
in {\it C81-04-06.1.4}
Phys.\ Rev.\ D {\bf 24}, 1808 (1981).


\bibitem{Braaten:1987yy}
E.~Braaten and S.~Tse,
Phys.\ Rev.\ D {\bf 35}, 2255 (1987).

\bibitem{Gronberg:1998fj}
J.~Gronberg {\em et al.} [CLEO Collaboration],
{Phys. Rev.} {\bf D57}, 33 (1998),  \hfill\break
hep-ex/9707031; and H. Paar, presented at PHOTON 2000: International
Workshop on Structure and Interactions of the Photon  Ambleside, Lake
District, England, 26-31 Aug 2000.


\bibitem{Radyushkin:1995pj}
A.~V.~Radyushkin,
Acta Phys.\ Polon.\ {\bf B26}, 2067 (1995)
[hep-ph/9511272].

\bibitem{Ong:1995gs}
S.~Ong,
Phys.\ Rev.\ D {\bf 52}, 3111 (1995).

\bibitem{Kroll:1996jx}
P.~Kroll and M.~Raulfs,
Phys.\ Lett.\ {\bf B387}, 848 (1996)
[hep-ph/9605264].

\bibitem{Chernyak:1984ej}
V.~L.~Chernyak and A.~R.~Zhitnitsky,
Phys.\ Rept.\ {\bf 112}, 173 (1984).

\bibitem{Farrar:1990qj}
G.~R.~Farrar and H.~Zhang,
Phys.\ Rev.\ Lett.\ {\bf 65}, 1721 (1990).

\bibitem{Brooks:2000nb}
T.~C.~Brooks and L.~Dixon,
Phys.\ Rev.\ D {\bf 62}, 114021 (2000)
[hep-ph/0004143].

\bibitem{Isgur:1989cy}
N.~Isgur and C.~H.~Llewellyn Smith,
Phys.\ Lett.\ {\bf B217}, 535 (1989).

\bibitem{Radyushkin:1998rt}
A.~V.~Radyushkin,
Phys.\ Rev.\ D {\bf 58}, 114008 (1998)
[hep-ph/9803316].

\bibitem{Bolz:1996sw}
J.~Bolz and P.~Kroll,
Z.\ Phys.\ {\bf A356}, 327 (1996)
[hep-ph/9603289].

\bibitem{Vogt:2000bz}
C.~Vogt,
hep-ph/0010040.

\bibitem{Paar}
Paar, H.,  {\em et al. } CLEO collaboration (to be published); See also
Boyer, J. \etal, {\em Phys.\ Rev.\ Lett.} {\bf 56}, 207 (1980);
TPC/Two Gamma Collaboration (H. Aihara \etal), {\em Phys. Rev. Lett.} {\bf
57},404 (1986).


\bibitem{Anderson:1973cc}
R.~L.~Anderson {\it et al.},
Phys.\ Rev.\ Lett.\ {\bf 30}, 627 (1973).

\bibitem{Sommermann:1992yh}
H.~M.~Sommermann, R.~Seki, S.~Larson and S.~E.~Koonin,
Phys.\ Rev.\ D {\bf 45}, 4303 (1992).

\bibitem{Marek:2001}
S. Brodsky and M. Karliner, in preparation.


\bibitem{BHDP}
S.~Brodsky, ~M.~Diehl, ~P.~Hoyer, and ~S.~Peigne, in preparation.

\bibitem{Aitala:2000hc}
E.~M.~Aitala {\it et al.}  [E791 Collaboration],
hep-ex/0010044.


\bibitem{Aitala:2000hb}
E.~M.~Aitala {\it et al.}  [E791 Collaboration],
pion light-cone wave function squared,''
hep-ex/0010043.



\bibitem{Stoler:1999nj}
P.~Stoler,
Few Body Syst.\ Suppl.\ {\bf 11}, 124 (1999).




\bibitem{Miller:1999mi}
G.~A.~Miller,
nucl-th/9910053.

\bibitem{Miller:2000ta}
G.~A.~Miller, S.~J.~Brodsky and M.~Karliner,
Phys.\ Lett.\ {\bf B481}, 245 (2000)
[hep-ph/0002156].


\bibitem{Franz:2000ee}
M.~Franz, M.~V.~Polyakov and K.~Goeke,
Phys.\ Rev.\ D {\bf 62}, 074024 (2000)
[hep-ph/0002240].

\bibitem{Brodsky:1981se}
S.~J.~Brodsky, C.~Peterson and N.~Sakai,
Phys.\ Rev.\ D {\bf 23}, 2745 (1981).


\bibitem{Chang:2001iy}
C.~V.~Chang and W.~Hou,
in $B$ Meson,''
hep-ph/0101162.


\bibitem{Brodsky:1990db}
S.~J.~Brodsky and I.~A.~Schmidt,
Phys.\ Lett.\ {\bf B234}, 144 (1990).


\bibitem{Brodsky:1997fj}
S.~J.~Brodsky and M.~Karliner,
Phys.\ Rev.\ Lett.\ {\bf 78}, 4682 (1997)
[hep-ph/9704379].

\bibitem{Brodsky:1987bb}
S.~J.~Brodsky, G.~P.~Lepage and S.~F.~Tuan,
Phys.\ Rev.\ Lett.\ {\bf 59}, 621 (1987).

\bibitem{Burkardt:1992di}
M.~Burkardt and B.~Warr,
Phys.\ Rev.\ D {\bf 45}, 958 (1992).



\bibitem{Brodsky:1996hc}
S.~J.~Brodsky and B.~Ma,
Phys.\ Lett.\ {\bf B381}, 317 (1996)
[hep-ph/9604393].

\bibitem{Pauli:1985ps}
H.~C.~Pauli and S.~J.~Brodsky,
Phys.\ Rev.\ D {\bf 32}, 2001 (1985).

\bibitem{Dalley:1993yy}
S.~Dalley and I.~R.~Klebanov,
Phys.\ Rev.\ D {\bf 47}, 2517 (1993)
[hep-th/9209049].



\bibitem{Antonuccio:1996hv}
F.~Antonuccio and S.~Dalley,
Phys.\ Lett.\ {\bf B376}, 154 (1996)
[hep-th/9512106].

\bibitem{Antonuccio:1996rb}
F.~Antonuccio and S.~Dalley,
Nucl.\ Phys.\ {\bf B461}, 275 (1996)
[hep-ph/9506456].

\bibitem{Brodsky:1998hs}
S.~J.~Brodsky, J.~R.~Hiller and G.~McCartor,
Phys.\ Rev.\ D {\bf 58}, 025005 (1998)
[hep-th/9802120].

\bibitem{Brodsky:1999xj}
S.~J.~Brodsky, J.~R.~Hiller and G.~McCartor,
Phys.\ Rev.\ D {\bf 60}, 054506 (1999)
[hep-ph/9903388].

\bibitem{Dalley:2000ii}
S.~Dalley and B.~van de Sande,
Phys.\ Rev.\ D {\bf 62}, 014507 (2000)
[hep-lat/9911035].



\bibitem{Hiller:1999cv}
J.~R.~Hiller and S.~J.~Brodsky,
Phys.\ Rev.\ D {\bf 59}, 016006 (1999)
[hep-ph/9806541].

\bibitem{Antonuccio:1999ia}
F.~Antonuccio, I.~Filippov, P.~Haney, O.~Lunin, S.~Pinsky,
U.~Trittmann and J.~Hiller [SDLCQ Collaboration],
hep-th/9910012.
Brodsky:1999xj
\bibitem{Lunin:1999ib}
O.~Lunin and S.~Pinsky,
hep-th/9910222.

\bibitem{Haney:2000tk}
P.~Haney, J.~R.~Hiller, O.~Lunin, S.~Pinsky and U.~Trittmann,
Phys.\ Rev.\ D {\bf 62}, 075002 (2000)
[hep-th/9911243].

\bibitem{Lepage:1982gd}
G.~P.~Lepage, S.~J.~Brodsky, T.~Huang and P.~B.~Mackenzie,
CLNS-82/522, published
in Banff Summer Inst.1981:0083 (QCD161:B23:1981);
S.~J.~Brodsky, T.~Huang
and G.~P.~Lepage,
{\it  In *Banff 1981, Proceedings, Particles and Fields 2*, 143-199}.

\bibitem{Daniel:1991ah}
D.~Daniel, R.~Gupta and D.~G.~Richards,
Phys.\ Rev.\ D {\bf 43}, 3715 (1991).

\bibitem{DelDebbio:2000mq}
L.~Del Debbio, M.~Di Pierro,
A.~Dougall and C.~Sachrajda  [UKQCD collaboration],
Nucl.\ Phys.\ Proc.\ Suppl.\ {\bf 83}, 235 (2000)
[hep-lat/9909147].


\bibitem{Dalley:2000dh}
S.~Dalley,
hep-ph/0007081.

\bibitem{Bardeen:1976tm}
W.~A.~Bardeen and R.~B.~Pearson,
Phys.\ Rev.\ D {\bf 14}, 547 (1976).

\bibitem{Burkardt:1996gp}
M.~Burkardt,
Phys.\ Rev.\ D {\bf 54}, 2913 (1996)
[hep-ph/9601289].

\bibitem{Petrov:1999kg}
V.~Y.~Petrov, M.~V.~Polyakov, R.~Ruskov, C.~Weiss and K.~Goeke,
Phys.\ Rev.\ D {\bf 59}, 114018 (1999)
[hep-ph/9807229].

\bibitem{Diakonov:2000pa}
D.~Diakonov and V.~Y.~Petrov,
hep-ph/0009006.


\end{thebibliography}
\end{document}